\documentclass[]{spie}  %>>> use for US letter paper
\usepackage[]{graphicx}
\usepackage{amsmath}
\usepackage{enumerate}
\usepackage{color}
\usepackage{cite}
\usepackage{enumitem}
\usepackage{mathptmx}      % use Times fonts if available on your TeX system
\usepackage{bm}% bold math symbols \bm\alpha

\title{Event-by-event simulation of experiments to create entanglement and violate Bell inequalities}
\author{K. Michielsen\supit{a,}\supit{b} and H. De Raedt\supit{c}
\skiplinehalf
\supit{a}
Institute for Advanced Simulation, J\"ulich Supercomputing Centre,
Forschungszentrum J\"ulich, D-52425 J\"ulich, Germany
\\
\supit{b}
RWTH Aachen University, D-52056 Aachen, Germany
\\
\supit{c}
Department of Applied Physics,
Zernike Institute for Advanced Materials,
University of Groningen, Nijenborgh 4, NL-9747 AG Groningen, The Netherlands
}

\authorinfo{
Further author information: (Send correspondence to K. Michielsen)\\
K. Michielsen: E-mail: k.michielsen@fz-juelich.de\\
H. De Raedt : E-mail: h.a.de.raedt@rug.nl
}
\newcommand{\url}[1]{{\rm #1}}

\begin{document}
  \maketitle

%%%%%%%%%%%%%%%%%%%%%%%%%%%%%%%%%%%%%%%%%%%%%%%%%%%%%%%%%%%%%
\begin{abstract}
We discuss a discrete-event, particle-based simulation approach which reproduces the statistical distributions
of Maxwell's theory and quantum theory by generating detection events one-by-one.
This event-based approach gives a unified cause-and-effect description of quantum optics
experiments such as single-photon Mach-Zehnder interferometer, Wheeler's delayed choice,
quantum eraser, double-slit, Einstein-Podolsky-Rosen-Bohm and
Hanbury Brown-Twiss experiments, and various neutron interferometry
experiments
at a level of detail which is not covered by conventional quantum theoretical descriptions.
We illustrate the approach by application to single-photon Einstein-Podolsky-Rosen-Bohm experiments
and single-neutron interferometry experiments that violate a Bell inequality.
\end{abstract}

%>>>> Include a list of keywords after the abstract


\section{Introduction}
In quantum theory entanglement is the property of a state of a two or many-body quantum system in which the constituting bodies are correlated.
The entangled state plays a prominent role in a thought experiment, devised in 1935 by
Einstein, Podolsky and Rosen (EPR) to demonstrate the ``incompleteness'' of quantum theory.~\cite{EPR35}
The thought experiment involves the measurement of the position and momentum of two particles which interacted in the past but not at the
time of measurement. Since this experiment is not suited for designing a laboratory experiment, Bohm proposed in 1951 a more realistic
experiment which measures the intrinsic angular momentum of a correlated pair of atoms one-by-one~\cite{BOHM51}.
Many experimental realizations and quantum theoretical descriptions of the EPR thought experiment~\cite{EPR35} adopt this model by Bohm,
which from now on we also refer to as experiment I.

The experiment consists of a particle source and two measurement stations each consisting of a Stern Gerlach magnet with two detectors
placed behind it.
The source emits charge-neutral pairs of particles with opposite magnetic moments $+{\mathbf S}$ and $-{\mathbf S}$.
Note that nothing is known about the direction of ${\mathbf S}$ itself.
The two particles separate spatially. One of the particles moves in free space to measurement station 1 positioned on the left hand side of
the source and the other moves in free space to station 2 positioned on the right hand side of the source.
As the particle arrives at station $j=1,2$, it passes through a Stern-Gerlach magnet. The magnetic moment of the particle interacts
with the inhomogeneous magnetic field of the Stern-Gerlach magnet.
The Stern-Gerlach magnet deflects the particle, depending on the orientation of the magnet ${\mathbf a}_j$ and the magnetic moment of the particle.
The Stern-Gerlach magnet divides the beam of particles in two, spatially well-separated parts. As the particle leaves the Stern-Gerlach magnet,
it generates a signal in one of the two detectors ${\mathrm D}_{\pm ,j}$. The firing of a detector corresponds to a detection event.
Coincidence logic pairs the detection events of station 1 and station 2 so that they can be used to compute two-particle correlations.

According to quantum theory of the Einstein-Podolsky-Rosen-Bohm (EPRB) thought experiment, the results of repeated
measurements of the system of two spin-1/2 particles in the spin state
$|\Psi\rangle =c_0 \left|\uparrow\uparrow\rangle\right. +c_1 \left|\downarrow\uparrow\rangle\right. +c_2 \left|\uparrow\downarrow\rangle\right. +c_3 \left|\downarrow\downarrow\rangle\right.$
with $\sum_{j=0}^3|c_j|^2=1$ are given by the single-spin expectation values
\begin{eqnarray}
\widehat E_1({\mathbf a}_1) &=& \langle\Psi |{\mathbf \sigma}_1 \cdot {\mathbf a}_1 |\Psi\rangle = \langle\Psi |{\mathbf \sigma}_1  |\Psi\rangle \cdot {\mathbf a}_1,\nonumber \\
\widehat E_2({\mathbf a}_2) &=& \langle\Psi |{\mathbf \sigma}_2 \cdot {\mathbf a}_2 |\Psi\rangle = \langle\Psi |{\mathbf \sigma}_2  |\Psi\rangle \cdot {\mathbf a}_2,
\label{twospins}
\end{eqnarray}
and the two-particle correlations $\widehat E({\mathbf a}_1,{\mathbf a}_2)=\langle\Psi |{\mathbf \sigma}_1 \cdot {\mathbf a}_1 {\mathbf \sigma}_2 \cdot {\mathbf a}_2 |\Psi\rangle
={\mathbf a}_1\cdot\langle\Psi |{\mathbf \sigma}_1\cdot {\mathbf \sigma}_2 |\Psi\rangle \cdot {\mathbf a}_2$, where
${\mathbf a}_1$ and ${\mathbf a}_2$ are unit vectors specifying the directions of the analyzers,
${\mathbf \sigma}_i$ denote the Pauli vectors describing the spin of the particles $j=1,2$, and
$\langle X\rangle ={\mathrm Tr}\rho X$ with $\rho$ being the 4x4 density matrix describing the two spin-1/2 particle system.
We have introduced the notation $\widehat {\phantom{E}}$ to make a distinction between the quantum theoretical results
and the results obtained from experiment (see Sect.~2.1) or an event-based simulation (see Sect.~2.2).
Quantum theory of the EPRB thought experiment assumes that $|\Psi\rangle$ does not depend
on ${\mathbf a}_1$ or ${\mathbf a}_2$. Therefore, from Eq.~(\ref{twospins}) it follows immediately that
$\widehat E_1({\mathbf a}_1)$ does not depend
on ${\mathbf a}_2$ and that $\widehat E_2({\mathbf a}_2)$ does not depend on ${\mathbf a}_1$.
Note that this holds for any state $|\Psi\rangle$.
For later use, it is expedient to introduce the function
\begin{equation}
\widehat S\equiv \widehat S({\mathbf a}_1,{\mathbf a}_2,{\mathbf a}_1^{\prime},{\mathbf a}_2^{\prime})=\widehat E({\mathbf a}_1,{\mathbf a}_2)-
\widehat E({\mathbf a}_1,{\mathbf a}_2^{\prime})+\widehat E({\mathbf a}_1^{\prime},{\mathbf a}_2)+ \widehat E({\mathbf a}_1^{\prime},{\mathbf a}_2^{\prime}),
\label{Bellfunction}
\end{equation}
for which it can be shown that $|\widehat S|\le 2\sqrt{2}$, independent of the choice of $\rho$.~\cite{CIRE80}
The function $\widehat S$ is often used to test the Bell-CHSH (Clauser-Holt-Shimony-Horne) inequality~\cite{CLAU69} $|\widehat S|\le 2$.

The quantum theoretical description of the EPRB experiment (experiment I) assumes that the state of the two spin-1/2 particles is described by
the singlet state $\rho=|\Psi\rangle\langle\Psi|$ where
\begin{equation}
|\Psi\rangle = \frac {1}{\sqrt{2}}\left(\left|\uparrow\downarrow\rangle\right. - \left|\downarrow\uparrow\rangle\right.\right).
\label{singlet}
\end{equation}
For the singlet state, $\widehat E_1({\mathbf a}_1)=\widehat E_2({\mathbf a}_2)=0$,
$\widehat E({\mathbf a}_1,{\mathbf a}_2)=-{\mathbf a}_1\cdot {\mathbf a}_2\equiv -\cos \alpha_{12}$,
the correlation $\widehat\rho_{12} ({\mathbf a}_1,{\mathbf a}_2)=\widehat E({\mathbf a}_1,{\mathbf a}_2)-
\widehat E_1({\mathbf a}_1)\widehat E_2({\mathbf a}_2)=\widehat E({\mathbf a}_1,{\mathbf a}_2)$
and the maximum value of $|\widehat S|$ is $2\sqrt{2}$.
Note that the singlet state is fully characterized by the three quantities $\widehat E_1({\mathbf a}_1)$, $\widehat E_2({\mathbf a}_2)=0$,
and $\widehat E({\mathbf a}_1,{\mathbf a}_2)$.
Hence, in any laboratory experiment, thought experiment or computer simulation of such an experiment, which has the goal to measure effects of the system being represented
by a singlet state, these three quantities have to be measured and computed, respectively.

We now discuss some variations of the EPRB thought experiment.
Experiment II is performed in the same way as experiment I but the particle source is replaced by a source which is emitting particles having
definite magnetic moments ${\mathbf S}_j$ for $j=1,2$.
The quantum theoretical description of experiment II assumes that the state of the two spin-1/2 particles is described by
the uncorrelated quantum state $\rho=\rho_1\bigotimes\rho_2$ where $\rho_j=|\theta_j\phi_j\rangle\langle\theta_j\phi_j|$
is the $2\times 2$ density matrix of particle $j$ and
\begin{equation}
|\theta_j\phi_j\rangle = \cos (\theta_j/2) \left|\uparrow\rangle\right. +e^{i\phi_j}\sin (\theta_j/2) \left|\downarrow\rangle\right.,
\label{product}
\end{equation}
for $j=1,2$.
For the uncorrelated quantum state, $\widehat E_j({\mathbf a}_j)={\mathbf a}_j\cdot {\mathbf S}_j\equiv\cos\alpha_j$ for $j=1,2$,
where ${\mathbf S}_j =(\cos\phi_j\sin\theta_j,\sin\phi_j\sin\theta_j,\cos\theta_j)$,
$\widehat E({\mathbf a}_1,{\mathbf a}_2)=\widehat E_1({\mathbf a}_1)\widehat E_2({\mathbf a}_2)=
({\mathbf a}_1\cdot {\mathbf S}_1)({\mathbf a}_2\cdot {\mathbf S}_2)=\cos\alpha_1\cos\alpha_2$,
$\widehat \rho_{12}({\mathbf a}_1,{\mathbf a}_2)=0$
and the maximum value of $|\widehat S|$ is 2.

Experiment III is performed in the same way as experiment I but between the source and measurement station $j$ a device producing
a uniform magnetic field with orientation $\mathbf{\eta}_j$ is placed. This device changes the magnetic moment $\pm \mathbf{S}$,
with unknown orientation, of the emitted particle into a magnetic moment with definite orientation $\mathbf{\eta}_j$.
Hence, the quantum theoretical description of this experiment assumes that the state of the two spin-1/2 particles
is described by an uncorrelated quantum state, just as in experiment II.

The results for the single and two particle expectation values and the correlations in the three experiments are summarized in Table~\ref{tab1}.
Within the framework of quantum theory the Bell-CHSH inequality $|\widehat S|<2$ can be used to make a distinction between the outcome of experiment I and experiments II and III.
If the state of the two spin-1/2 particle system is an uncorrelated quantum state, then the
Bell-CHSH inequality holds. On the other hand, if the Bell-CHSH
inequality is violated then the two-particle quantum system is in a correlated (entangled) state.

Several so-called Bell test experiments have been performed to find two-particle correlations which correspond to those of the singlet state.
In this paper we discuss two of them, namely a single-photon EPRB experiment claiming that the two photons of a pair,
post-selected by employing a time-coincidence window, can be in an entangled state~\cite{WEIH98,WEIH00} and
a neutron interferometry experiment~\cite{HASE03} which shows that it is possible to create correlations
between the spatial and spin degree of freedom of
neutrons which, within quantum theory, cannot be described
by a product state meaning that the spin- and
phase-degree-of-freedom are entangled.
In the latter experiment
the neutrons are counted with a detector having a
very high efficiency ($\approx 99\%$), thereby not suffering from
the so-called detection loophole.

We will demonstrate that the event-based simulation method,~\cite{MICH11a,RAED12a,RAED12b} which uses simple rules
to define discrete-event-processes, simulates the behavior that is observed in the single-photon and neutron
interferometry Bell test experiments
and in the related experiments II and III of the single-photon experiment.
The method is entirely classical
in the sense that it uses concepts of the macroscopic
world and makes no reference to quantum theory but is
nonclassical in the sense that some of the rules are not
those of classical Newtonian dynamics.

\begin{table*}%[H] add [H] placement to break table across pages
\caption{\label{tab1}
Single and two-particle expectation values for a quantum system of two spin-1/2 particles in the
singlet state and the uncorrelated quantum state.
}
%\begin{ruledtabular}
\centering
\begin{tabular}{lcc}
\noalign{\medskip}
\hline\hline\noalign{\smallskip}
 &  Singlet state & Uncorrelated quantum state \\
\hline\noalign{\smallskip}
    $\widehat E_1({\mathbf a}_1)$ &    0&    ${\mathbf a}_1\cdot {\mathbf S}_1\equiv\cos\alpha_1$   \\
    $\widehat E_2({\mathbf a}_2)$ &    0 &   ${\mathbf a}_2\cdot {\mathbf S}_2\equiv\cos\alpha_2$   \\
    $\widehat E({\mathbf a}_1,{\mathbf a}_2)$ &    $-{\mathbf a}_1\cdot {\mathbf a}_2\equiv -\cos\alpha_{12}$ & $({\mathbf a}_1\cdot {\mathbf S}_1)({\mathbf a}_2\cdot {\mathbf S}_2)=\cos\alpha_1\cos\alpha_2$    \\
    $\widehat \rho_{12}({\mathbf a}_1,{\mathbf a}_2)$ &   $ -\cos\alpha_{12}$ &    0   \\
\hline\noalign{\smallskip}
\end{tabular}
%\end{ruledtabular}
\end{table*}

\section{EPRB and modified experiments with single photons}
In the single-photon experiments, the polarization of each photon plays the role of the spin-1/2 degree-of-freedom in Bohm's version~\cite{BOHM51}
of the EPR thought experiment~\cite{EPR35}.
Using the fact that the two-dimensional vector space with basis vectors $\{|H\rangle,|V\rangle\}$,
where $H$ and $V$ denote the horizontal and vertical polarization of the photon, respectively, is isomorphic to the vector space
with basis vectors $\{\left|\uparrow\rangle\right. ,\left|\downarrow\rangle\right.\}$ of spin-1/2 particles, we may use the
language of the latter to describe the experiments I, II and III with photons.
For photons the antisymmetric (singlet) state reads
\begin{equation}
|\Psi\rangle=\frac {1}{\sqrt{2}}\left(|H\rangle_1|V\rangle_2 - |V\rangle_1|H\rangle_2\right)
=\frac {1}{\sqrt{2}}\left(|HV\rangle - |VH\rangle\right),
\label{singletphoton}
\end{equation}
and the uncorrelated quantum state reads
\begin{equation}
|\Psi\rangle=\left(\cos\zeta_1|H\rangle_1+\sin\zeta_1|V\rangle_1\right)\left(\cos\zeta_2|H\rangle_2+\sin\zeta_2|V\rangle_2\right),
\label{productphoton}
\end{equation}
where $\zeta_j$ for $j=1,2$ denotes the definite polarization of the photons and the subscripts refer to photon 1 and 2, respectively.
The polarization vector ${\mathbf P}_j=(\cos\zeta_j,\sin\zeta_j,0)$ replaces the magnetic moment ${\mathbf S}_j =(\cos\phi_j\sin\theta_j,\sin\phi_j\sin\theta_j,\cos\theta_j)$ of the spin-1/2 particle.
The expressions for the single-photon expectation values and the two-photon correlations are similar to those of the
genuine spin-1/2 particle problem except for the restriction of $\mathbf{a}_1$ and $\mathbf{a}_2$ to lie in planes
orthogonal to the direction of propagation of the photons and that the polarization is defined modulo $\pi$, not modulo $2\pi$ as
in the case of the spin-1/2 particles. The latter results in a multiplication of the angles by a factor of two.
For simplicity it is often assumed that ${\mathbf a}_j=(\cos a_j,\sin a_j,0)$ for $j=1,2$.
The resulting single and two particle expectation values and the correlations are summarized in Table~\ref{tab2}.

\begin{table*}%[H] add [H] placement to break table across pages
\caption{\label{tab2}
Single and two-particle expectation values for a quantum system of two photons in the
singlet state and the uncorrelated quantum state.
}
%\begin{ruledtabular}
\centering
\begin{tabular}{lcc}
\noalign{\medskip}
\hline\hline\noalign{\smallskip}
 &  Singlet state & Uncorrelated quantum state \\
\hline\noalign{\smallskip}
    $\widehat E_1({\mathbf a}_1)$ &    0&    ${\mathbf a}_1\cdot {\mathbf S}_1\equiv\cos 2\alpha_1=\cos 2(\zeta_1-a_1)$   \\
    $\widehat E_2({\mathbf a}_2)$ &    0 &   ${\mathbf a}_2\cdot {\mathbf S}_2\equiv\cos 2\alpha_2=\cos 2(\zeta_2-a_2)$   \\
    $\widehat E({\mathbf a}_1,{\mathbf a}_2)$ &    $ -\cos 2\alpha_{12}=-\cos 2(a_1-a_2)$ &
    $\cos 2\alpha_1\cos 2\alpha_2=\cos 2(\zeta_1-a_1)\cos 2(\zeta_2-a_2)$    \\
    $\widehat \rho_{12}({\mathbf a}_1,{\mathbf a}_2)$ &   $ -\cos 2\alpha_{12}=-\cos 2(a_1-a_2)$ &    0   \\
\hline\noalign{\smallskip}
\end{tabular}
%\end{ruledtabular}
\end{table*}

\begin{figure}[pt]
\begin{center}
\includegraphics[width=12cm]{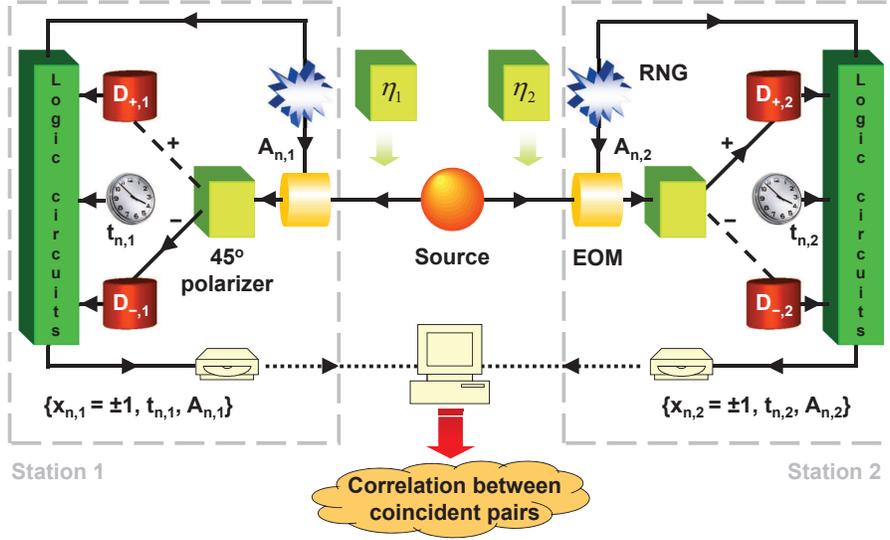}
\vspace*{8pt}
\caption{%
Schematic diagram of experiments I~\cite{WEIH98,WEIH00}, II and III with single photons.
The source emits pairs of photons.
One of the photons moves to station 1 and the other one to station 2.
In experiment I, the photons have orthogonal but otherwise random polarization.
In experiment II, the photons have orthogonal but definite polarization ($\eta_2=\eta_1+\pi/2$).
In experiment III the source is the same as in experiment I
but a polarizer with definite orientation ${\mathbf \eta}_j =(\cos\eta_j,\sin\eta_j,0)$
is placed between the source and measurement station $j$. The polarizer changes the indefinite polarization of the emitted
photon into the definite polarization ${\mathbf \eta}_j$.
As the photon arrives at station $j=1,2$ it first passes through an electro-optic modulator (EOM) which
rotates the polarization of the photon by an angle $\varphi_j$ depending on the voltage applied to the EOM.
This voltage is controlled by a binary variable $A_j$, which is chosen at random.
As the photon leaves the EOM, a polarizing beam splitter directs it to one of the two detectors ${\mathrm D}_{\pm ,j}$.
The detector produces a signal $x_{n,j}=\pm 1$ where the subscript $n$ labels the $n$th detection event.
Each station has its own clock which assigns a time-tag $t_{n,j}$ to each detection signal.
A data set $\left\{ {x_{n,j},t_{n,j},A_{n,j} \vert n=1,\ldots ,N_j } \right\}$ is stored on a hard disk for each station.
Long after the experiment is finished both data sets can be analyzed and among other things, two-particle correlations can be computed.
}
\label{fig1}
\end{center}
\end{figure}

\subsection{Laboratory experiment}
We take the EPRB experiment with single photons, which corresponds to experiment I, carried out by Weihs {\sl et al.}~\cite{WEIH98,WEIH00},
as a concrete example.
We first describe the data collection and analysis procedure of the experiment.
Next we describe how this experiment can be modified to study experiments II and III.
Then we illustrate how to construct an event-based model of an idealized version of these experiments which
reproduces the predictions of quantum theory for the single and two-particle averages
for a quantum system of two spin-1/2 particles in the singlet state and
a product state~\cite{ZHAO08,MICH11a}, without making reference to concepts of quantum theory.

\begin{enumerate}[leftmargin=*]
\item{{\sl Data collection:}
Figure~\ref{fig1} shows a schematic diagram of the EPRB experiment with single photons carried out by Weihs {\sl et al.}~\cite{WEIH98,WEIH00}
(experiment I).
The source emits pairs of photons with orthogonal but otherwise random polarization.
The photon pair splits and each photon travels in free space to
an observation station, labeled by $j=1$ or $j=2$, in which it is manipulated and
detected. The two stations are assumed to be identical and are separated spatially and temporally.
Hence, the observation at station 1 (2) cannot have a causal effect on the data registered at station 2 (1).~\cite{WEIH98}
As the photon arrives at station $j=1,2$ it first passes through an electro-optic modulator (EOM) which
rotates the polarization of the photon by an angle $\varphi_j$ depending on the voltage applied to the EOM.~\cite{WEIH98,WEIH00}
This voltage is controlled by a binary variable $A_j$, which is chosen at random.~\cite{WEIH98,WEIH00}
Optionally, a bias voltage is added to the randomly varying voltage.~\cite{WEIH98,WEIH00}
The relation between the voltage applied to the EOM and the resulting rotation of the polarization
is determined experimentally, hence there is some uncertainty in relating the applied voltage
to the rotation angle.~\cite{WEIH98,WEIH00}
As the photon leaves the EOM, a polarizing beam splitter directs it to one of the two detectors.
The detector produces a signal $x_{n,j}=\pm1$ where the subscript $n$ labels the $n$th detection event.
Each station has its own clock which assigns a time-tag $t_{n,j}$ to each signal generated by one of the two detectors.~\cite{WEIH98,WEIH00}
Effectively, this procedure discretizes time in intervals, the width of which is
determined by the time-tag resolution $\tau$.
In the experiment, the time-tag generators are synchronized before each run.~\cite{WEIH98,WEIH00}

The firing of a detector is regarded as an event.
At the $n$th event at station $j$,
the dichotomic variable $A_{n,j}$,
controlling the rotation angle $\varphi_{n,j}$,
the dichotomic variable $x_{n,j}$ designating which detector fires,
and the time tag $t_{n,j}$ of the detection event
are written to a file on a hard disk,
allowing the data to be analyzed long after the experiment has terminated.~\cite{WEIH98,WEIH00}
The set of data collected at station $j$ may be written as
\begin{eqnarray}
\label{Ups}
\Upsilon_j=\left\{ {x_{n,j},t_{n,j},\varphi_{n,j} \vert n =1,\ldots ,N_j } \right\}
,
\label{eprb1}
\end{eqnarray}
where we allow for the possibility that the number of detected events $N_j$
at stations $i=1,2$ need not (and in practice is not) to be the same
and we have used the rotation angle $\varphi_{n,j}$ instead
of the corresponding experimentally relevant dichotomic variable $A_{n,j}$ to facilitate the
comparison with the quantum theoretical description.
}%
\item{{\sl Data analysis procedure:}
A laboratory EPRB experiment requires some criterion to decide which detection
events are to be considered as stemming from a single or two-particle system.
In EPRB experiments with photons, this decision is taken on the basis of coincidence in time.~\cite{WEIH98,CLAU74}
Here we adopt the procedure employed by Weihs {\sl et al.}~\cite{WEIH98,WEIH00}
Coincidences are identified by comparing the time differences
$t_{n,1}-t_{m,2}$ with a window $W$,~\cite{WEIH98,WEIH00,CLAU74}
where $n=1,\ldots,N_1$ and $m=1,\ldots,N_2$.
By definition, for each pair of rotation angles $a_1$ and $a_2$,
the number of coincidences between detectors $D_{x,1}$ ($x =\pm $1) at station 1 and
detectors $D_{y,2}$ ($y =\pm $1) at station 2 is given by
\begin{eqnarray}
\label{Cxy}
C_{xy}&=&C_{xy}(a_1,a_2) \nonumber \\
&=&
\sum_{n=1}^{N_1}
\sum_{m=1}^{N_2}
\delta_{x,x_{n ,1}} \delta_{y,x_{m ,2}}
\delta_{a_1 ,\varphi_{n,1}}\delta_{a_2,\varphi_{m,2}}
%\nonumber \\&&\times
\Theta(W-\vert t_{n,1} -t_{m ,2}\vert)
,
\end{eqnarray}
where $\Theta (t)$ denotes the unit step function.
In Eq.~(\ref{Cxy}) the sum over all events has to be carried out such that each event (= one detected photon) contributes only once.
Clearly, this constraint introduces some ambiguity in the counting procedure as there is a priori, no clear-cut criterion
to decide which events at stations $j=1$ and $j=2$ should be paired.
One obvious criterion might be to choose the pairs such that $C_{xy}$ is maximum, but
such a criterion renders the data analysis procedure (not the data production) acausal.
It is trivial though to analyze the data generated by the experiment of Weihs {\sl et al.}
such that conclusions do not suffer from this artifact.~\cite{RAED12}
In general, the values for the coincidences
$C_{xy}(a_1,a_2)$ depend on the time-tag resolution $\tau$
and the window $W$ used to identify the coincidences.

The single-particle averages and correlation between the coincidence counts
are defined by
\begin{eqnarray}
\label{Exy}
E_1(a_1,a_2)&=&
\frac{\sum_{x,y=\pm1} x C_{xy}}{\sum_{x,y=\pm1} C_{xy}}
=
\frac{C_{++}-C_{--}+C_{+-}-C_{-+}}{N_c}
\nonumber \\
E_2(a_1,a_2)&=&
\frac{\sum_{x,y=\pm1} yC_{xy}}{\sum_{x,y=\pm1} C_{xy}}
=
\frac{C_{++}-C_{--}-C_{+-}+C_{-+}}{N_c}
\nonumber \\
E(a_1,a_2)&=&
\frac{\sum_{x,y=\pm1} xy C_{xy}}{\sum_{x,y=\pm1} C_{xy}}
%\nonumber \\&=&
=
\frac{C_{++}+C_{--}-C_{+-}-C_{-+}}{N_c}
,
\end{eqnarray}
where the denominator $N_c=N_c(a_1,a_2)=C_{++}+C_{--}+C_{+-}+C_{-+}$
in Eq.~(\ref{Exy}) is the sum of all coincidences.

Local-realistic treatments of the EPRB experiment assume that the correlation,
as measured in the experiment, is given by~\cite{BELL93}
\begin{eqnarray}
\label{CxyBell}
C_{xy}^{(\infty)}(a_1,a_2)&=&\sum_{n=1}^N\delta_{x,x_{n ,1}} \delta_{y,x_{n ,2}}
\delta_{a_1,\theta_{n,1}}\delta_{a_2,\theta_{m,2}}
,
\end{eqnarray}
which is obtained from Eq.~(\ref{Cxy})
(in which each photon contributes only once)
by assuming that $N=N_1=N_2$, pairs are defined by $n=m$ and
by taking the limit $W\rightarrow\infty$.
However, the working hypothesis that the value of $W$ should not matter
because the time window only serves to identify pairs may not apply to real experiments.
The analysis of the data of the experiment of Weihs {\sl et al.} shows that
the average time between pairs of photons is of the order of $30\mu$s or more,
much larger than the typical values (of the order of a few nanoseconds)
of the time-window $W$ used in the experiments.~\cite{WEIH00}
In other words, in practice, the identification of photon pairs
does not require the use of $W$'s of the order of a few nanoseconds.

An analysis of in total 23 data sets produced by the experiment of Weihs {\sl et al.} shows that
none of these data sets satisfies the hypothesis that the statistics of this data is described by quantum theory.~\cite{RAED12}
Although the experiment generates data that violate Bell inequalities for suitable choices of the time-coincidence window,
it is also shown that for the same choices of the time-coincidence window $E_1(a_1)$ depends on $a_2$ and that $E_2(a_2)$ depends $a_1$, making it highly
unlikely that the data is compatible with quantum theory of two photons.
In another paper in this volume we demonstrate that the EPRB experiments of M.B. Ag\"uero {et al.}~\cite{AGUE09} and Adenier {\sl et al.}~\cite{VIST12,ADEN12} show the same features.
This suggests that the conclusion that single photon experiments agree with quantum theory is premature and that more
precise experiments are called for.
}%
\item{{\sl Modification for experiment II and III:}
To perform experiment II the single-photon source is replaced by a source which emits photons with orthogonal but definite
polarization, $\eta_2=\eta_1+\pi /2$ (see Fig.~\ref{fig1}). To perform experiment III a polarizer with definite orientation ${\mathbf \eta}_j =(\cos\eta_j,\sin\eta_j,0)$
is placed between the source and measurement station $j$ (see Fig.~\ref{fig1}).
The polarizer changes the indefinite polarization of the emitted
photon into the definite polarization ${\mathbf \eta}_j$.
}%
\end{enumerate}

\subsection{Event-based simulation}
A minimal, discrete-event simulation model of the EPRB experiment by Weihs {\sl et al.} (experiment I) and of experiments II and III (see Fig.~\ref{fig1})
requires a specification of the information carried by the particles,
of the algorithm that simulates the source, the polarizers, the detectors, and of the procedure to analyze the data.
Since in the above description of the experiment the orientation of the polarization vectors ${\mathbf P}_j=(\cos\zeta_j,\sin\zeta_j,0)$ and
the orientations of the optical axis of the polarizers ${\mathbf a}_j=(\cos a_j,\sin a_j,0)$ for $j=1,2$ is
limited to the $xy$-plane we omit the $z$-component in the simulation.

\begin{enumerate}[leftmargin=*]
\item{{\sl Source and particles:}
Each time, the source emits two particles which carry a vector
${\mathbf u}_{n,j}=(\cos(\xi_{n}+(j-1)\pi/2) ,\sin(\xi_{n}+(j-1)\pi/2))$,
representing the polarization of the photons.
This polarization is completely characterized by the angle $\xi _{n}$ and
the direction $j=1,2$ to which the particle moves.
In case of experiment I, a uniform pseudo-random number generator is used to pick the angle $0\le\xi _{n}<2\pi$.
Clearly, the source emits two particles with a mutually orthogonal, hence correlated but otherwise
random polarization.
In case of experiment II, a predefined angle $0\le\xi_n=\xi <2\pi$ is used to represent the definite polarization of the photons.
Thus, in this case, the source emits two particles with a mutually orthogonal, definite
polarization.
%Note that for the simulation of this experiment it is not necessary that the particles carry information about the phase $2\pi ft_{i,n}$, although it would be possible.
%In this case the time of flight $t_{i,n}$ is determined by the time-tag model (see below).
}
\item{{\sl Electro-optic modulator} (EOM){\sl :}
The EOM in station $j=1,2$ rotates the polarization of the incoming particle by an angle $\varphi_j$,
that is its polarization angle becomes $\xi^{\prime}_{n,j}\equiv\mathrm{EOM}_j(\xi_{n}+(j-1)\pi/2,\varphi_j)=\xi_{n}+(j-1)\pi/2-\varphi_j$ symbolically.
Mimicking the experiment of Weihs {\sl et al.} in which $\varphi_1$ can take the values $a_1,a_1^{\prime}$ and $\varphi_2$ can take the values $a_2,a_2^{\prime}$,
we generate two binary uniform pseudo-random numbers $A_j=0,1$ and use them
to choose the value of the angles $\varphi_j$, that is
$\varphi_1=a_1(1-A_1)+a_1^{\prime}A_1$ and $\varphi_2=a_2(1-A_2)+a_2^{\prime}A_2$.
}
\item{{\sl Beam-splitting polarizer:}
In laboratory EPRB experiments with photons the
various polarizers are interchangeable. Therefore, the algorithm to simulate them should be identical.
Evidently, this should also hold for the polarizers placed in between the
source and the observation stations in experiment III.

%The input-output relation of a polarizer is rather simple. For each input event, the
%algorithm maps the input vector ${\mathbf u}$  onto a single output bit $x$. The value of the
%output bit depends on the orientation of the polarizer ${\mathbf a} = (\cos a, \sin a)$. According
%to Malus law, for ¯fixed ${\mathbf u} = (\cos\xi, \sin\xi)$ and fixed ${\mathbf a}$,
%the bits $x_n$ are to be generated such that
%
%\begin{equation}
%\lim_{N\rightarrow\infty}\frac{1}{N}\sum_{n=1}^Nx_n=\cos 2 (\xi-\alpha),
%\label{Malus}
%\end{equation}
%
%with probability one.
%If the input vectors ${\mathbf u}$ are distributed uniformly over the unit
%circle, the sequence of output bits should satisfy
%
%\begin{equation}
%\lim_{N\rightarrow\infty}\frac{1}{N}\sum_{n=1}^Nx_n=0,
%\end{equation}
%
%with probability one, independent of the orientation ${\mathbf \varphi}$ of the polarizer.

The simulation model for a beam-splitting polarizer is defined by the rule
\begin{eqnarray}
x_{n,j}=\left\{
\begin{array}{lll}
+1 & \mbox{if} & r_n\le \cos^2(\xi^{\prime}_{n,j})\\
-1 & \mbox{if} & r_n > \cos^2(\xi^{\prime}_{n,j})
\end{array}
\right.
,
\label{sg1}
\end{eqnarray}
where $0< r_n<1$ are uniform pseudo-random numbers.
The polarizer sends a photon with polarization ${\mathbf u}_n=(\cos \varphi_j,\sin \varphi_j)$ or
${\mathbf u}_n=(-\sin \varphi_j,\cos \varphi_j)$ through its output channel labeled by $+1$ and $-1$, respectively.
It is easy to see that for fixed $\xi^{\prime}_{n,i}=\xi^{\prime}_i$, this rule generates
events such that
%
%\begin{eqnarray}
$\lim_{N\rightarrow\infty} \sum_{n=1}^N x_{n,j}/N = \cos^2\varphi_{n,j}$
,
%\label{sg2}
%\end{eqnarray}
%
with probability one, showing that
the distribution of events complies with Malus law.
In experiment III we discard particles with polarization $\eta_1+\pi /2$ ($\eta_2 +\pi /2)$ that leave the
polarizers, placed in between the source and observation station 1 (2), via the output channel labeled by $-1$.

Note that this simplified mathematical model suffices to simulate the EPRB experiment but cannot be used to simulate all optics experiments
with beam-splitting polarizers
(for instance  Wheeler's delayed choice experiment).~\cite{MICH11a}
However, the more complicated models used to simulate the beam-splitting polarizer in these other experiments can be used to simulate the EPRB experiment.~\cite{MICH11a}
}
\item{{\sl Time-tag model:} %
As is well-known, as light passes through an EOM (which is essentially a tuneable wave plate), it experiences a retardation
depending on its initial polarization and the rotation by the EOM.
However, to our knowledge, time delays caused by retardation properties of waveplates, being components of various
optical apparatuses, have not yet been explicitly measured for single photons.
Therefore, in the case of single-particle experiments, we hypothesize that for each particle this delay is
represented by the time tag~\cite{RAED07b,ZHAO08}
%
%\begin{eqnarray}
$t_{n,i}=\lambda(\xi^{\prime}_{n,i}) r^{\prime}_n$,
%,
%\label{sg3b}
%\end{eqnarray}
which is distributed uniformly ($0<r^{\prime}_n <1$ is a uniform pseudo-random number) over the interval $[0, \lambda(\xi^{\prime}_{n,i})]$.
For $\lambda(\xi^{\prime}_{n,i})=T_0\sin^4 2\xi^{\prime}_{n,i}$ this time-tag model, in combination with the model
of the polarizing beam splitter, rigorously reproduces the results
of quantum theory of the EPRB experiments in the limit $W\rightarrow0$~\cite{RAED07b,ZHAO08}.
We therefore adopt the expression $\lambda(\xi^{\prime}_{n,i})=T_0 \sin^4 2\xi^{\prime}_{n,i}$
leaving only $T_0$ as an adjustable parameter.
}
\item{{\sl Detector:} %
The detectors are ideal particle counters, producing a click for each incoming particle.
Hence, we assume that the detectors have 100\% detection efficiency,
which makes the data collecting procedure free from the detection loophole.
Simulating adaptive threshold detectors is a trivial modification and does not change our main conclusions.~\cite{MICH11a}
}
\item{{\sl Simulation and data analysis procedure:} %
The simulation algorithm generates the data sets $\Upsilon_i$, similar to the ones obtained in the experiment (see Eq.~(\ref{eprb1})).
In the simulation, it is easy to generate the events such that $N_1=N_2$.
We analyze these data sets in exactly the same manner as the experimental data are analyzed, implying that we
include the post-selection procedure to select photon pairs by a time-coincidence window $W$.
In order to count the coincidences, we choose a time-tag resolution $0 < \tau < T_0$ and a coincidence window $\tau \le W$. We set
the correlation counts $C_{xy}(\varphi_1,\varphi_2)$ to zero for all $x,y = \pm 1$.
We compute the discretized time tags
$k_{n,j} = \lceil t_{n,j}/\tau \rceil$ for all events in both data sets. Here $\lceil x\rceil $ denotes
the smallest integer that is larger or equal to $x$, that is
$\lceil x\rceil -1 < x \le \lceil x\rceil $. According to the procedure adopted in the
experiment~\cite{WEIH98,WEIH00}, an entangled photon pair is observed if and
only if $|k_{n,1}-k_{n,2}| < k = \lceil W/\tau\rceil$. Thus, if $|k_{n,1}-k_{n,2}| < k$, we
increment the count $C_{x_{n,1},x_{n,2}} (\varphi_1, \varphi_2)$.
%The post-selection procedure is crucial for our simulation method to give results that are very similar to those observed in a laboratory experiment.
Although in the simulation the ratio of detected to emitted photons is equal to one, the final detection efficiency is reduced
due to the time-coincidence post-selection procedure thereby introducing a time-coincidence loophole.
}
\end{enumerate}

\begin{figure}[pt]
\begin{center}
\includegraphics[width=8.5cm]{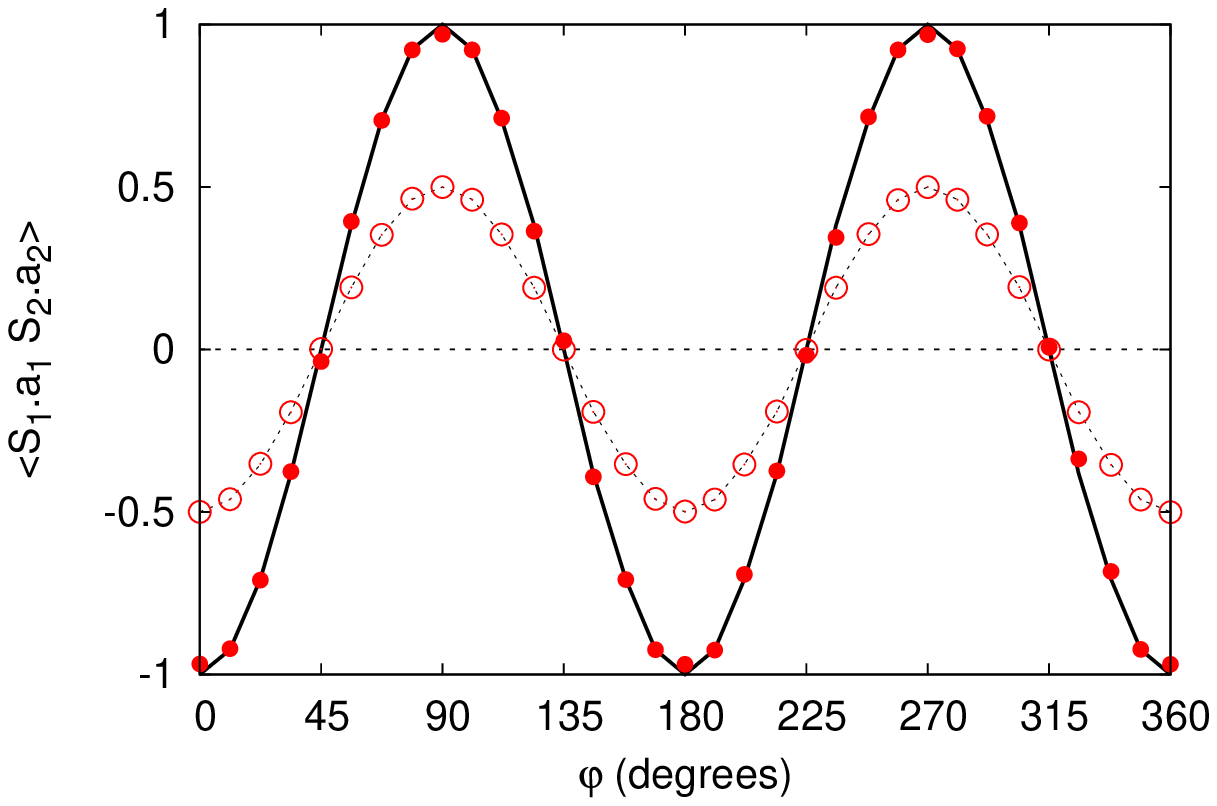}
\includegraphics[width=8.5cm]{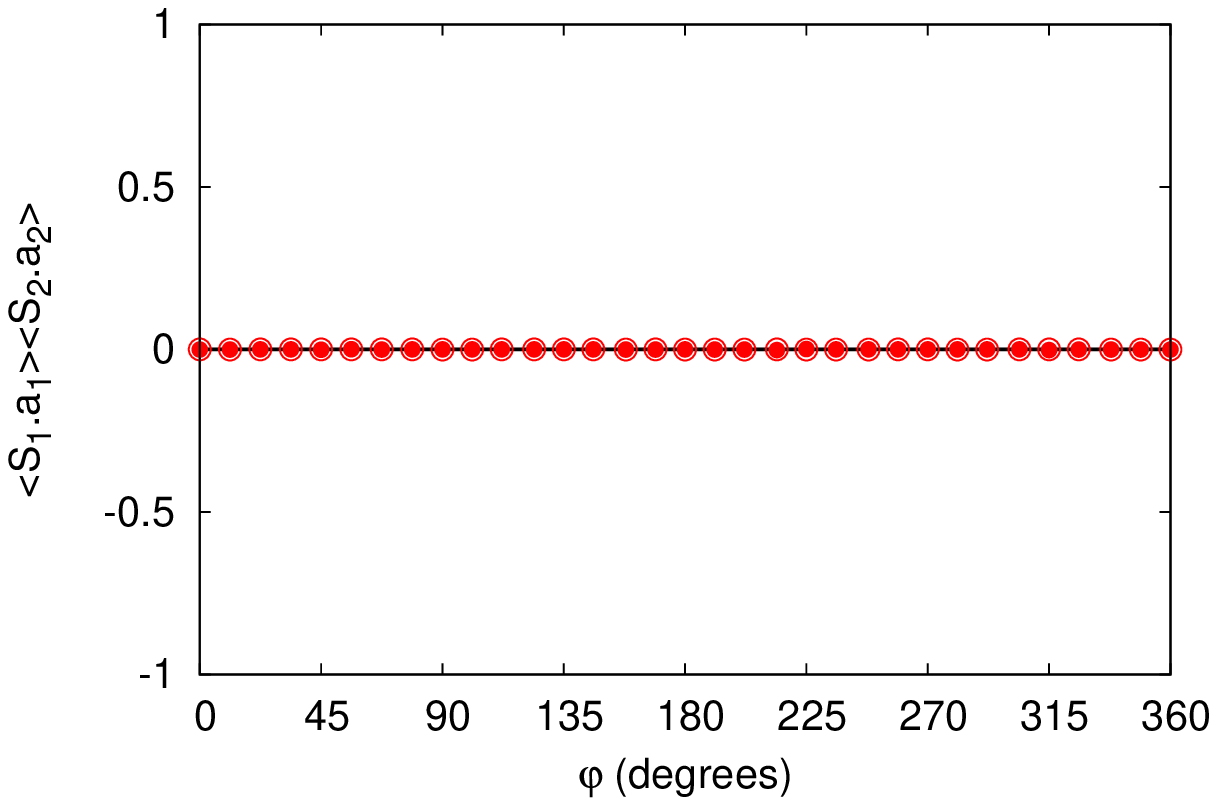}
\vspace*{8pt}
\caption{%
Simulation results (markers) and quantum theoretical result (solid line) for the EPRB experiment (experiment I) with $\varphi_1=\varphi$ and $\varphi_2=0$ for
the two-particle expectation value $\widehat E(\varphi)=\langle {\mathbf S}_1\cdot {\mathbf a}_1 {\mathbf S}_2\cdot {\mathbf a}_2\rangle$
(left) and the
product of the two single-particle expectation values $\widehat E_1(\varphi)\widehat E_2(\varphi)=\langle {\mathbf S}_1\cdot {\mathbf a}_1\rangle \langle{\mathbf S}_2\cdot {\mathbf a}_2\rangle$
(right) as a function of $\varphi$.
In experiment I, the source emits two photons with orthogonal but otherwise random polarization.
The number of emitted photon pairs $N=(N_1+N_2)/2=10^6$ with $N_1=N_2$ and the adjustable parameter in the time-tag model $T_0=10^3$.
Solid circles:  coincidence counting with $W/\tau=1$; open circles: no coincidence counting.
The dashed line through the open circles is a guide to the eye.}
\label{fig2}
\end{center}
\end{figure}

\subsection{Simulation results}
\begin{enumerate}[leftmargin=*]
\item{{\sl Experiment I:}
Figure \ref{fig2} presents the simulation results (markers) for experiment I with $\varphi_1=\varphi$ and $\varphi_2=0$ for
the two-particle expectation value $\widehat E(\varphi)=\langle {\mathbf S}_1\cdot {\mathbf a}_1 {\mathbf S}_2\cdot {\mathbf a}_2\rangle$
(left) and the
product of the two single-particle expectation values $\widehat E_1(\varphi)\widehat E_2(\varphi)=\langle {\mathbf S}_1\cdot {\mathbf a}_1\rangle \langle{\mathbf S}_2\cdot {\mathbf a}_2\rangle$
(right) as a function of $\varphi$.
The figure shows both the data resulting from a coincidence counting data analysis procedure (solid markers) as well as the data from a data analysis procedure without coincidence counting (open markers).
The results expected from the quantum theoretical description of experiment I are $\widehat E(\varphi,0)=-\cos 2\varphi$ and $\widehat E_1(\varphi)=\widehat E_2(0)=0$
and are represented by the solid lines.

The coincidence counting data analysis procedure with $W/\tau=1$ (solid markers), which is similar to the one used in the experiment by Weihs {\sl et al.}~\cite{WEIH98,WEIH00},
gives results which fit very well to the prediction of quantum theory for the EPRB experiment.
For relatively small time-coincidence windows $W/\tau$ (not all results shown), the single and two-particle expectation values of the singlet can be obtained and therefore the maximal
value of $|S|=2.82$ is obtained.

However, if all detected photons are taken into account (open markers), which corresponds to a data analysis procedure without using a time-coincidence window $W$ to select pairs,
then $E(\varphi,0)=-(\cos 2\varphi )/2=\widehat E(\varphi,0)/2$.
Note that this data analysis procedure is equivalent to a procedure in which $W\rightarrow\infty$ or to a procedure in which the time-tag data is simply omitted.

The difference between coincidence counting or not in the data analysis procedure clearly demonstrates the fact that the observation of two-particle correlations
$\rho_{12}(\varphi,0)$ corresponding to those of the singlet state
$\widehat\rho_{12}(\varphi,0)=-\cos 2\varphi$ strongly depends on how the data is measured (including time-tags of the detection events) and analyzed (size of the time-coincidence window).
This information does not exist in a simplistic ``singlet state'' description of how photon pairs are generated.
}
\begin{figure}[pt]
\begin{center}
\includegraphics[width=8.5cm]{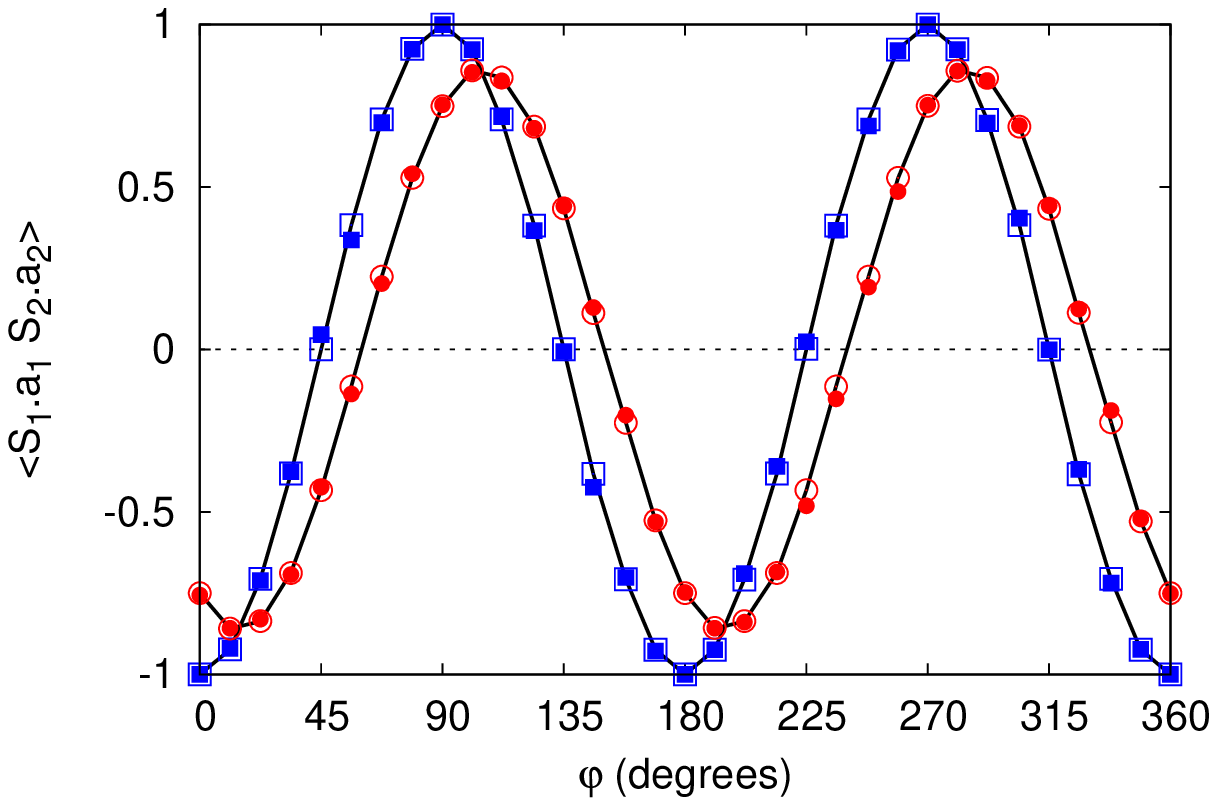}
\includegraphics[width=8.5cm]{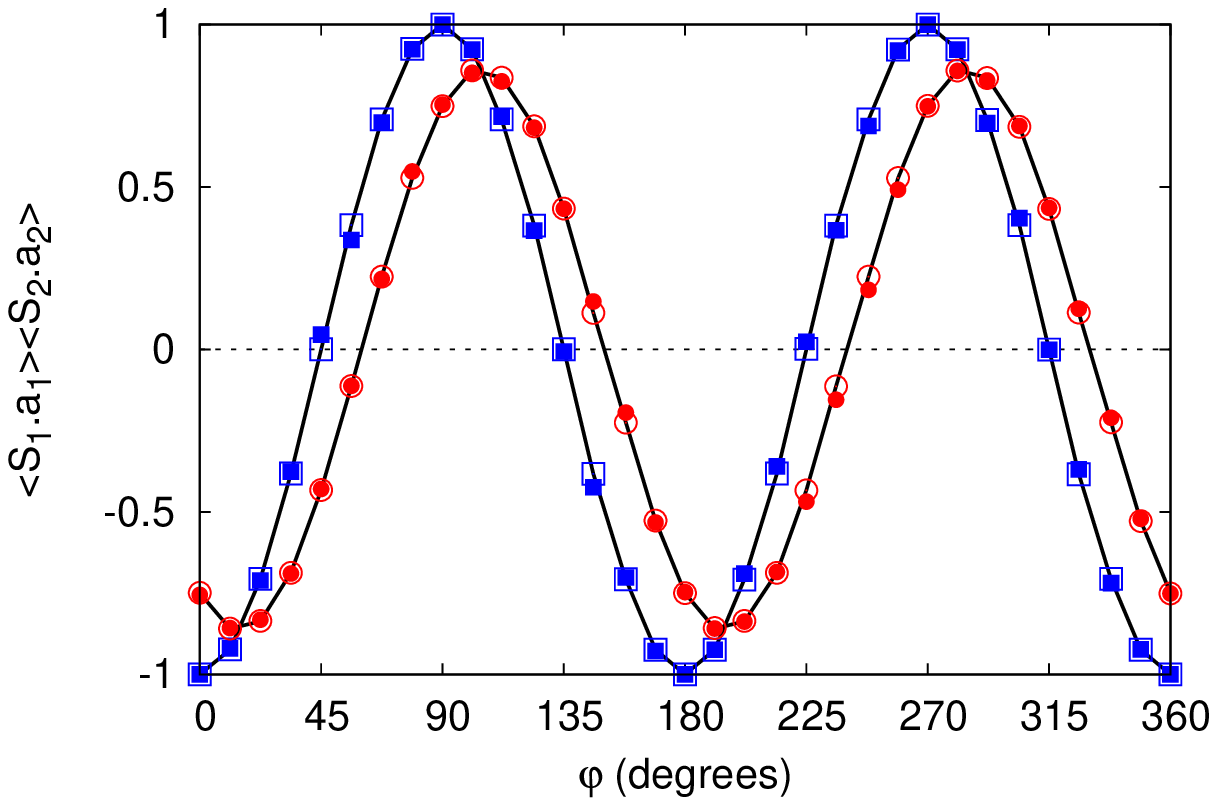}
\vspace*{8pt}
\caption{%
Same as Fig.~\ref{fig2} for experiment II.
In experiment II, the source emits two photons with orthogonal but definite polarization ($\eta_2=\eta_1+\pi/2$).
Blue squares: $\eta_1=0$, $\eta_2=90^\circ$; red circles: $\eta_1=15^\circ$, $\eta_2=105^\circ$.
}
\label{fig3}
\end{center}
\end{figure}
\item{{\sl Experiment II:}
Similar simulation results for experiment II with photons leaving the source with definite polarizations $\eta_2=\eta_1+\pi/2$ with $\eta_1 =0$, $\eta_2=90^\circ$ (blue squares)
and $\eta_1 =15^\circ$, $\eta_2=105^\circ$ (red circles), as for experiment I are depicted in Fig.~\ref{fig3}.
The quantum theoretical description of experiment II gives $\widehat E(\varphi,0)=\cos 2(\eta_1-\varphi)\cos 2\eta_2$, $\widehat E_1(\varphi)=\cos 2(\eta_1-\varphi)$ and $\widehat E_2(0)=\cos 2\eta_2$,
represented by the solid lines.
Both data analysis procedures, with or without coincidence counting, give results which fit very well to the quantum theoretical description of experiment II in terms of an uncorrelated quantum state.

Note that for $\eta_1=0$ and $\eta_2=90^\circ$,  $\widehat E(\varphi,0)=-\cos 2\varphi$, corresponding to the two-particle expectation value of a singlet state, but $\widehat \rho_{12}(\varphi ,0)=0$.
This demonstrates that it is essential to measure both the two-particle and one-particle expectation values in an experiment.
}
\begin{figure}[pt]
\begin{center}
\includegraphics[width=8.5cm]{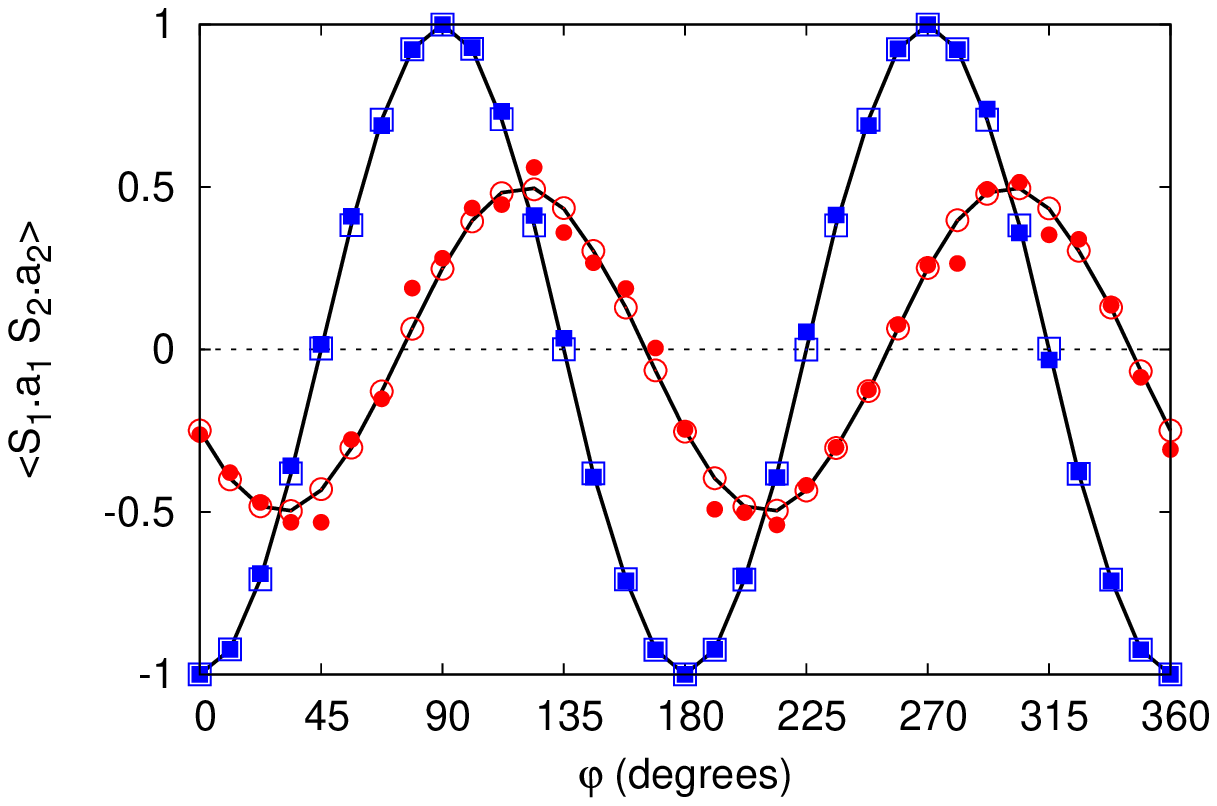}
\includegraphics[width=8.5cm]{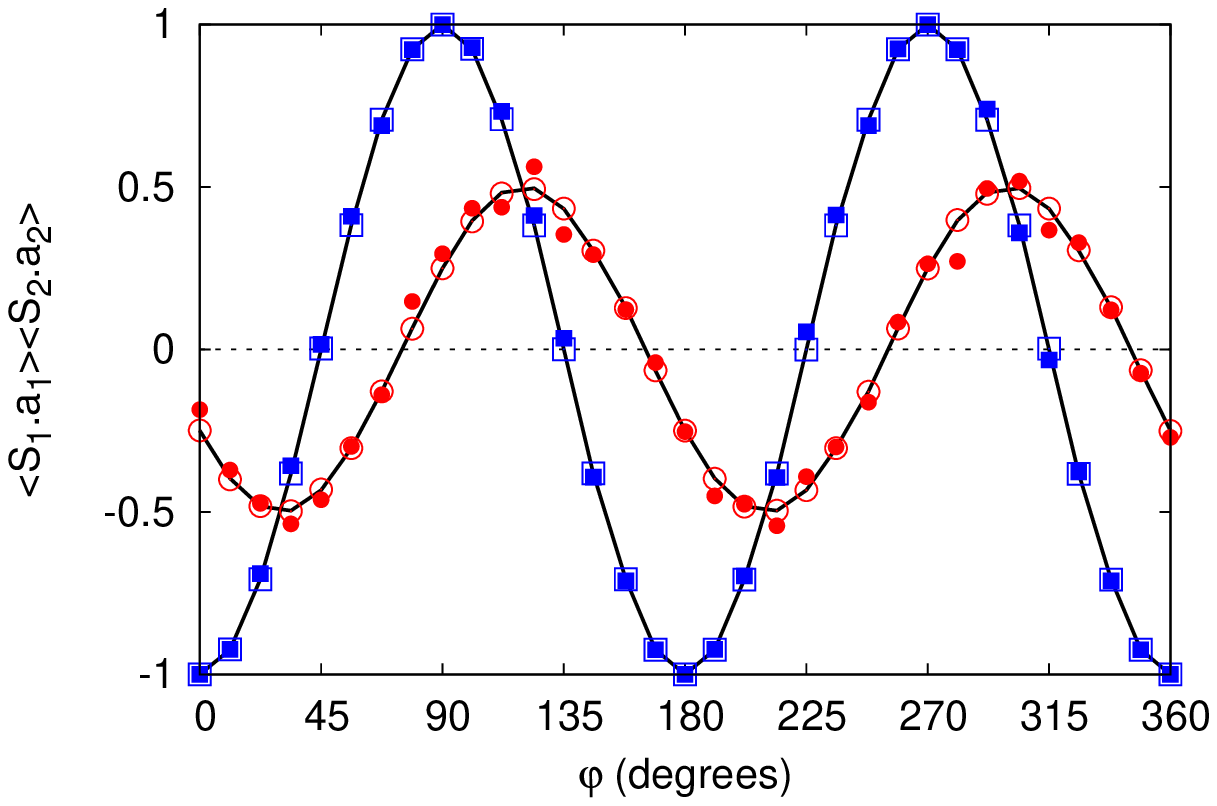}
\vspace*{8pt}
\caption{%
Same as Fig.~\ref{fig2} for experiment III.
In experiment III the source emits two photons with orthogonal but otherwise random polarization,
but a polarizer with definite orientation ${\mathbf \eta}_j$
is placed between the source and measurement station $j$ to change the indefinite polarization of the emitted
photon into the definite polarization ${\mathbf \eta}_j$ for $j=1,2$.
Blue squares: $\eta_1=0$, $\eta_2=90^\circ$; red circles: $\eta_1=30^\circ$, $\eta_2=60^\circ$.
}
\label{fig4}
\end{center}
\end{figure}
\item{{\sl Experiment III:}
Simulation results for experiment III are shown in Fig.~\ref{fig4}.
In this experiment the photons leaving the source have orthogonal but otherwise random polarization,
but a polarizer placed between the source and measurement station $j$ changes the indefinite polarization of the emitted
photon into the definite polarization ${\mathbf \eta}_j$ for $j=1,2$.
The quantum theoretical description of this experiment is the same as for experiment II.
The simulation results are in perfect agreement with the quantum theoretical description.
Examples for $\eta_1=30^\circ$, $\eta_2=60^\circ$ (red circles) and $\eta_1=0$, $\eta_2=90^\circ$ (blue squares) are presented.
}
\end{enumerate}

In summary, when $W\rightarrow0$ the discrete-event model which generates the same type
of data as a laboratory EPRB experiment (experiment I),
reproduces exactly the single- and two-spin averages of the singlet state and therefore also violates the
inequality $|S|\le2$.
Obviously, as the discrete-event model does not rely on any concept of quantum theory, a violation of the inequality $|S|\le2$ does not
say anything about the ``quantumness'' of the system under observation~\cite{KARL09,KARL10,RAED11a}.
Similarly, a violation of this inequality cannot say anything about locality and realism~\cite{KARL09,KARL10,RAED11a,NIEU11}.
Clearly, the event-based model is contextual, literally meaning ``being dependent of the (experimental) measurement arrangement''.

The fact that the event-based model reproduces, for instance, the correlations of the singlet state
without violating Einstein's local causality criterion suggests that the
data $\{x_{n,1},x_{n,2}\}$ generated by the event-based model cannot be represented by a single Kolmogorov probability space.
This complies with the idea that contextual, non-Kolmogorov models can lead to
violations of Bell's inequality without appealing to nonlocality or nonobjectivism~\cite{KHRE09,KHRE11}.

The same components to simulate the EPRB experiment (experiment I) can be used to simulate a quantum system of two polarized photons in an
uncorrelated quantum state (experiments II and III).
For experiments II and III the data analysis procedure with or without coincidence counting can be used to obtain results that are in correspondence
with the quantum theoretical description of the experiment. This is in contrast to experiment I for which only the data analysis procedure with the coincidence counting
gives the same results as the ones predicted by quantum theory.

\subsection{Why can Bell's inequality be violated?}
In Ref.~\citen{ZHAO08}, we have presented a probabilistic description of our simulation model that (i) rigorously proves that
for up to first order in $W$ it exactly reproduces the single particle averages and the two-particle correlations of
quantum theory for the system under consideration; (ii) illustrates how the presence of the time-window $W$ introduces
correlations that cannot be described by the original Bell-like ``hidden-variable'' models~\cite{BELL93}.
A discussion about the latter point is also presented in Ref.~\citen{RAED12a}.

Although the event-based simulation model involves local processes only, the filtering of the
detection events by means of the time-coincidence window $W$ can produce correlations which violate Bell-type
inequalities~\cite{FINE82,PASC86,LARS04}.
Moreover, for $W\rightarrow 0$ the classical (non-Hamiltonian like), local and causal simulation model can produce single-particle and two-particle
averages that correspond with those of a singlet state in quantum theory.
If the time-tag information ($W>T_0$) is ignored, the two-particle probability takes the form of the hidden variable models
considered by Bell~\cite{BELL93}, and the results of quantum theory cannot be reproduced.~\cite{BELL93}

\section{Bell-test experiment with single neutrons}
The single-neutron interferometry experiment of Hasegawa {\it et al.}~\cite{HASE03}
demonstrates that the correlation between
the spatial and spin degree of freedom of neutrons violates a Bell-CHSH inequality.
This Bell-test experiment thus involves two degrees of freedom of one particle, while the EPRB thought experiment~\cite{BOHM51}
and EPRB experiments with single photons~\cite{WEIH98,WEIH00,HNIL02,AGUE09} involve two degrees of freedom of two particles.
Hence, the single neutron Bell-test experiment is not performed according to the CHSH protocol~\cite{CLAU69} because the two degrees of freedom
of one particle are not manipulated and measured independently.
In this section we construct an event-based model that reproduces the correlation between the spatial and spin degree of freedom of the neutrons
by using detectors that count every neutron and without
using any post-selection procedure.

\begin{figure}[t]
\begin{center}
\includegraphics[width=9.5cm]{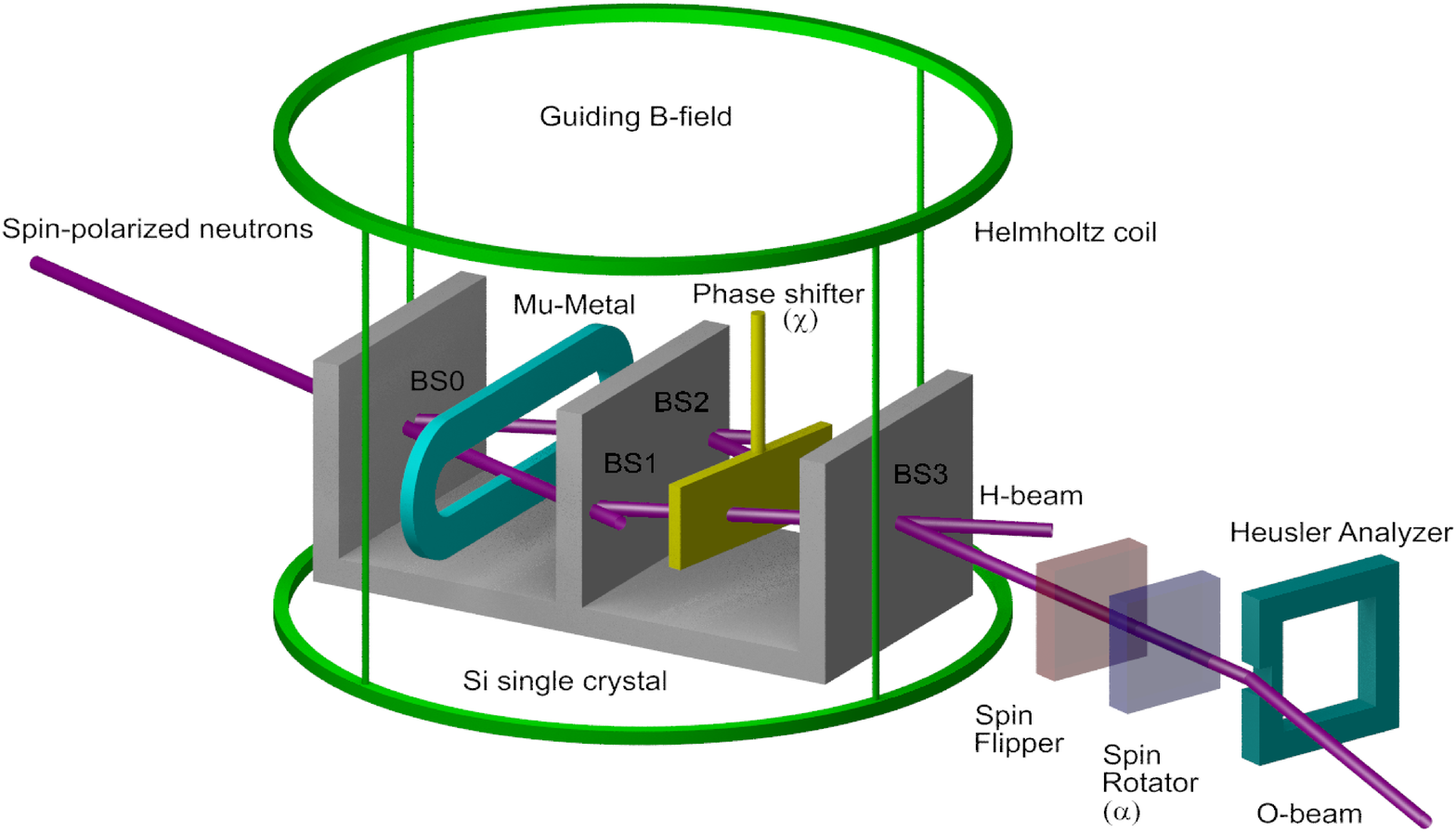}
\includegraphics[width=9.5cm]{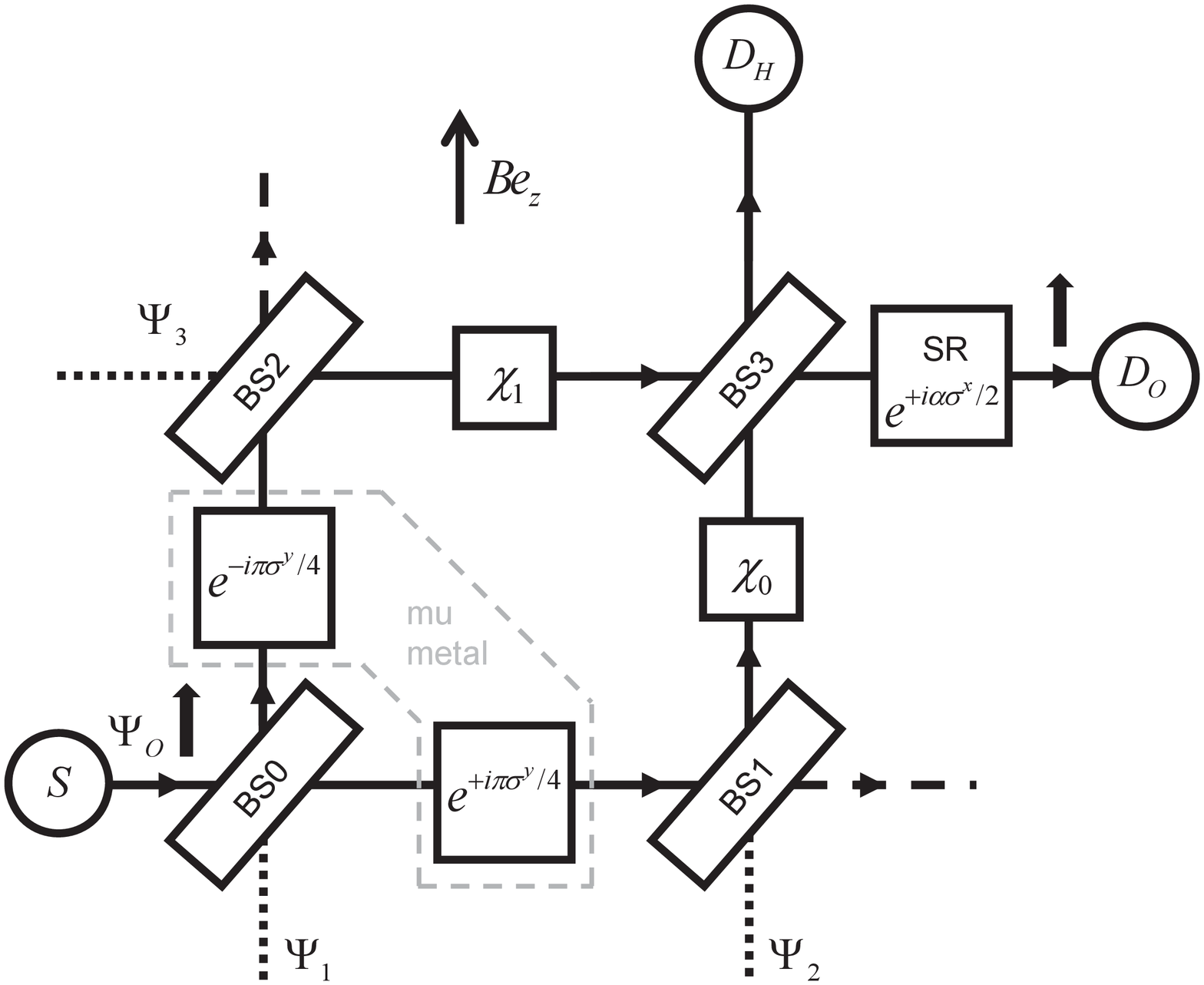}
\caption{%
Top: Schematic picture of the single-neutron interferometry experiment to test a Bell inequality violation (see also Fig.~1 in Ref.~\citen{HASE03}).
BS0, $\ldots$, BS3: beam splitters; phase shifter $\chi$: aluminum foil; neutrons that are transmitted by BS1 or BS2 leave the
interferometer and do not contribute to the interference signal. Detectors count the number of neutrons in the
O- and H-beam.
Bottom: Event-based network of the experimental setup shown on the top.
S: single neutron source; BS0, $\ldots$ , BS3: beam splitters;
$e^{+i\pi\sigma^y/4}$, $e^{-i\pi\sigma^y/4}$: spin rotators modeling the action of a mu metal;
$\chi_0$, $\chi_1$: phase shifters;
SR $e^{i\alpha\sigma^x/2}$: spin rotator;
$D_O$, $D_H$: detectors counting all neutrons
that leave the interferometer via the O- and H-beam, respectively. In the experiment and in
the event-based simulation, neutrons with spin up (magnetic moment aligned parallel with respect to the
guiding magnetic field ${\mathbf B}$) enter the interferometer via the path
labeled by $\Psi_0$ only. The wave amplitudes labeled by $\Psi_1$, $\Psi_2$, and $\Psi_3$ (dotted lines) are used in the quantum
theoretical treatment only. Particles leaving the interferometer via
the dashed lines are not counted.
}
\label{fig5}
\end{center}
\end{figure}

Figure~\ref{fig5} (top) shows a schematic picture of the single-neutron interferometry experiment.
Incident neutrons pass through a magnetic-prism polarizer (not shown) which produces two spatially separated beams of
neutrons with their magnetic moments aligned parallel (spin up), respectively anti-parallel (spin down) with respect
to the magnetic axis of the polarizer which is parallel to the guiding field ${\mathbf B}$. The spin-up neutrons
impinge on a silicon-perfect-crystal interferometer.~\cite{RAUC00} On leaving the first beam splitter BS0,
neutrons are transmitted or refracted.
A mu-metal spin-turner changes the orientation of the magnetic moment of the neutron from parallel to perpendicular to the guiding field ${\mathbf B}$.
Hence, the magnetic moment of the neutrons following path H (O) is rotated by $\pi/2$ ($-\pi/2$)
about the $y$ axis. Before the two paths join at the entrance plane of beam splitter BS3, a difference between the time of flights
along the two paths can be manipulated by a phase shifter. The neutrons
which experience two refraction events when passing through the interferometer form the O-beam and are analyzed by sending them through
a spin rotator and a Heusler spin analyzer. If necessary, to induce an extra spin rotation of $\pi$, a spin flipper is placed between
the interferometer and the spin rotator. The neutrons that are selected by the Heusler spin analyzer are counted with a
neutron detector (not shown) that has a very high efficiency ($\approx 99\%$).
Note that neutrons which are not refracted by the mirror plate leave the interferometer
without being detected.

The single-neutron interferometry experiment yields the count rate $N(\alpha,\chi)$ for the spin-rotation angle $\alpha$ and
the difference $\chi$ of the phase shifts of the two different paths in the interferometer~\cite{HASE03}.
The correlation $E(\alpha,\chi)$ is defined by~\cite{HASE03}

\begin{equation}
E(\alpha,\chi)=\frac{N(\alpha,\chi)+N(\alpha+\pi,\chi+\pi)-N(\alpha+\pi,\chi)-N(\alpha,\chi+\pi)}
{N(\alpha,\chi)+N(\alpha+\pi,\chi+\pi)+N(\alpha+\pi,\chi)+N(\alpha,\chi+\pi)}.
\end{equation}

\subsection{Event-based model}
A minimal, discrete event simulation model of the single-neutron interferometry experiment requires a specification of the
information carried by the particles, of the algorithm that simulates the source and the interferometer components
(see Fig.~\ref{fig5} (bottom)), and of the procedure to analyze the data.

\begin{itemize}[leftmargin=*]
\renewcommand\labelitemi{-}
\item{{\sl Source and particles:}
A neutron is regarded as a messenger carrying a
message represented by the two-dimensional unit vector
\begin{equation}
{\mathbf u}=(e^{i\psi^{(1)}}\cos (\theta/2), e^{i\psi^{(2)}}\sin (\theta/2)),
\label{neutron}
\end{equation}
where
$\psi^{(i)} =\nu t +\delta_i$, for $i=1,2$.
Here, $t$ specifies the time of flight of the neutron
and $\nu$ is an angular frequency which is characteristic for a neutron
that moves with a fixed velocity $v$.
A monochromatic
beam of incident neutrons is assumed to consist of neutrons that
all have the same value of $\nu$.~\cite{RAUC00}
Both $\theta$ and $\delta=\delta_1-\delta_2=\psi^{(1)}-\psi^{(2)}$ determine the magnetic moment of the neutron, if
the neutron is viewed as a tiny classical magnet spinning around the direction
${\mathbf m}=(\cos\delta\sin \theta, \sin\delta\sin \theta, \cos \theta)$,
relative to a fixed frame of reference defined by a magnetic field.
Note that this is only a pictorial description
of the mathematical representation and nothing more.
The third degree of freedom in Eq.~(\ref{neutron}) is used to account
for the time of flight of the neutron.
Within the present model, the state of the neutron is fully determined
by the angles $\psi^{(1)}$, $\psi^{(2)}$ and $\theta$ and by rules (to be specified), by which these angles change
as the neutron travels through the network.

A messenger with message ${\mathbf u}$ at time $t$ and position ${\mathbf r}$
that travels with velocity $v$, along the direction ${\mathbf q}$ during a time interval
$t^{\prime} - t$, changes its message according to $\psi^{(i)}\leftarrow \psi^{(i)}+\phi$ for $i=1,2$, where
$\phi =\nu(t^{\prime}-t)$.

In the presence
of a magnetic field ${\mathbf B}=(B_x,B_y,B_z)$, the magnetic moment
rotates about the direction of ${\mathbf B}$ according to the
classical equation of motion. Hence, in a magnetic field the message ${\mathbf u}$ is changed into the message
${\mathbf w}=e^{ig\mu_NT\mathbf {\sigma}\cdot {\mathbf B}/2} {\mathbf u}$, where
$g$ denotes the neutron $g$-factor, $\mu_N$ the nuclear magneton, $T$ the time during which the neutron experiences the magnetic field, and $\mathbf {\sigma}$ the vector of the three Pauli matrices
(here we use the isomorphism between the algebra of Pauli matrices and rotations in three-dimensional space).

When the source creates a messenger, its message needs to be initialized.
This means that the three angles $\psi^{(1)}$, $\psi^{(2)}$ and $\theta$ need to be specified.
The specification depends on the type of source that has to be simulated.
For a fully coherent spin-polarized beam of neutrons, the three angles are the same for all the messengers being created.
Hence, one random number is used to specify $\psi^{(1)}$, $\psi^{(2)}$ and $\theta$ for all messengers.
}
\item{{\sl Magnetic-prism polarizer:}
This component takes as input a neutron
with an unknown magnetic moment and produces a neutron
with a magnetic moment that is either parallel (spin up) or antiparallel
(spin down) with respect to the $z$-axis (which by definition
is parallel to the guiding field ${\mathbf B}$). In the experiment, only a
neutron with spin up is injected into the interferometer. Therefore,
as a matter of simplification, we assume that the source $S$ only creates messengers with spin up.
Hence, we assume that $\theta =0$ in Eq.~(\ref{neutron}).
}
\item{{\sl Beam splitters} BS0, $\ldots $ , BS3{\sl :}
A beam splitter is used to redirect neutrons depending on their magnetic moment.
In general, a beam splitter has two input and two output channels labeled by $k=0$ and $k=1$.
The beam splitter has two internal registers ${\mathbf R}_{k,n}=(R_{0,k,n},R_{1,k,n})$ with $R_{i,k,n}$ for $i=0,1$ representing a complex number, and an
internal vector ${\mathbf v}_n=(v_{0,n},v_{1,n})$, where $v_{i,n}\ge0$ for $i=0,1$, $v_{0,n}+v_{1,n}=1$
and $n$ denotes the message number.
The internal registers and the internal vector are labeled by the message number $n$ because their content is updated
for each messenger arriving at the beam splitter (see below).
Before the simulation starts uniform pseudo-random numbers are used to set
${\mathbf v}_0$, ${\mathbf R}_{0,0}$ and ${\mathbf R}_{1,0}$.

When the $n$th messenger carrying the message $\mathbf{u}_{k,n}$ arrives at entrance port $k=0$ or $k=1$ of the beam splitter,
the beam splitter first copies the message in the corresponding register ${\mathbf R}_{k,n}$ and updates its internal vector according to
\begin{equation}
{\mathbf v}_n=\gamma {\mathbf v}_{n-1}+(1-\gamma){\mathbf q}_n,
\label{internalPBS}
\end{equation}
where $0<\gamma<1$
and ${\mathbf q}_n=(1,0)$ (${\mathbf q}_n=(0,1)$) represents the arrival of the $n$th messenger on channel $k=0$ ($k=1$).
Note that storage is foreseen for exactly ten real-valued numbers.

Next the beam splitter uses the information stored in ${\mathbf R}_{0,n}$, ${\mathbf R}_{1,n}$ and ${\mathbf v}_n$ to
calculate four complex numbers
\begin{eqnarray}
\left(
\begin{array}{c}
h_{0,n}\\
h_{1,n}\\
h_{2,n}\\
h_{3,n}
\end{array}
\right)
&=&
\left(
\begin{array}{cccc}
\sqrt{{\cal T}}&i\sqrt{{\cal R}}&0&0\\
i\sqrt{{\cal R}}&\sqrt{{\cal T}}&0&0\\
0&0&\sqrt{{\cal T}}&i\sqrt{{\cal R}}\\
0&0&i\sqrt{{\cal R}}&\sqrt{{\cal T}}
\end{array}
\right)
\left(
\begin{array}{cccc}
\sqrt{v_{0,n}}&0&0&0\\
0&\sqrt{v_{1,n}}&0&0\\
0&0&\sqrt{v_{0,n}}&0\\
0&0&0&\sqrt{v_{1,n}}
\end{array}
\right)
\left(
\begin{array}{c}
R_{0,0,n}\\
R_{0,1,n}\\
R_{1,0,n}\\
R_{1,1,n}
\end{array}
\right)\nonumber \\
&=&
\left(
\begin{array}{c}
\sqrt{v_{0,n}}\sqrt{{\cal T}}R_{0,0,n}+i\sqrt{v_{1,n}}\sqrt{{\cal R}}R_{0,1,n}\\
i\sqrt{v_{0,n}}\sqrt{{\cal R}}R_{0,0,n}+\sqrt{v_{1,n}}\sqrt{{\cal T}}R_{0,1,n}\\
\sqrt{v_{0,n}}\sqrt{{\cal T}}R_{1,0,n}+i\sqrt{v_{1,n}}\sqrt{{\cal R}}R_{1,1,n}\\
i\sqrt{v_{0,n}}\sqrt{{\cal R}}R_{1,0,n}+\sqrt{v_{0,n}}\sqrt{{\cal T}}R_{1,1,n}
\end{array}
\right)
,
\label{PBS1neutron}
\end{eqnarray}
where the reflection $\cal{R}$ and transmission ${\cal T}=1-\cal{R}$ are real numbers
which are considered to be parameters to be determined from experiment,
and generates a uniform random number $r_n$ between
zero and one.
If $|h_{0,n}|^2+|h_{2,n}|^2 > r_n$, the beam splitter sends a message
%
%\begin{equation}
${\mathbf w}_{0,n}=(h_{0,n},h_{2,n})/\sqrt{|h_{0,n}|^2+|h_{2,n}|^2}$,
%\end{equation}
%
through output channel 1. Otherwise
it sends a message
%
%\begin{equation}
${\mathbf w}_{1,n}=(h_{1,n},h_{3,n})/\sqrt{|h_{1,n}|^2+|h_{3,n}|^2}$,
%\end{equation}
%
through output channel 0.
}
\item{{\sl Phase shifter $\chi_0$, $\chi_1$:}
%In the event-based model, a phase shifter is simulated without DLM.
The device has only one input and one output port and
transforms the $n$th input message ${\mathbf u}_n$ into an output message
%\begin {equation}
${\mathbf w}_n=e^{i\chi_j}{\mathbf u}_n$ for $j=0,1$.
%\end{equation}
}
\item{{\sl Mu metal spin turner:}
This component rotates the magnetic moment of a neutron that follows the H-beam (O-beam) by $\pi/2$ ($-\pi/2$) about the $y$ axis.
The processor that accomplishes this
takes as input the direction of the magnetic moment,
represented by the message $\mathbf u$ and performs
the rotation ${\mathbf u}\leftarrow e^{\pm i\pi\sigma ^y/4}{\mathbf u}$.
We emphasize that we use Pauli matrices as a convenient tool
to express rotations in three-dimensional space, not because in quantum theory
the magnetic moment of the neutron is represented by spin-1/2
operators.
}
\item{{\sl Spin-rotator and spin-flipper:}
The spin-rotator rotates the magnetic moment of a neutron by an angle $\alpha$ about the $x$ axis.
The spin flipper is a spin rotator with $\alpha=\pi$.
}
\item{{\sl Spin analyzer:}
This component selects neutrons with spin up, after which they are counted by a detector.
The model of this component projects the magnetic moment of the particle on the $z$ axis and sends the particle to the
detector if the projected value exceeds a pseudo-random number $r$.
}
\item{{\sl Detector:}
Detectors count all incoming particles.
Hence, we assume that the neutron detectors have a detection efficiency of 100\%.
This is an idealization of real neutron detectors which can have a detection effieciency of $99\%$ and more.~\cite{KROU00}
}
\end{itemize}
\begin{figure}[t]
\begin{center}
\includegraphics[width=8.5cm]{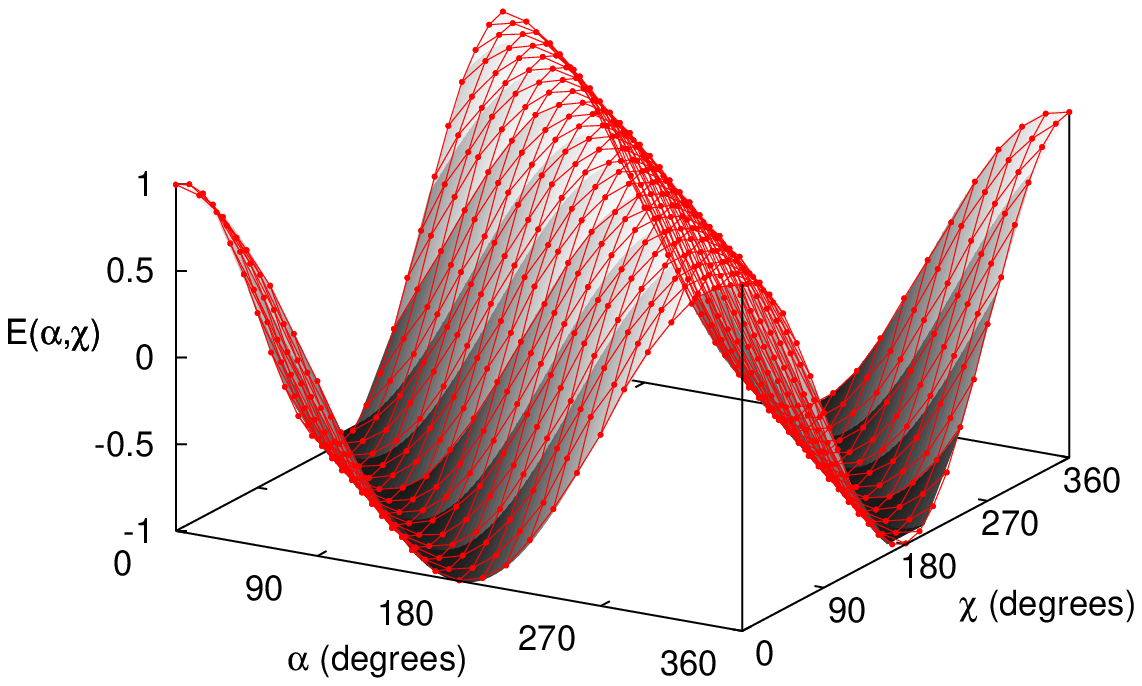}
\includegraphics[width=8.5cm]{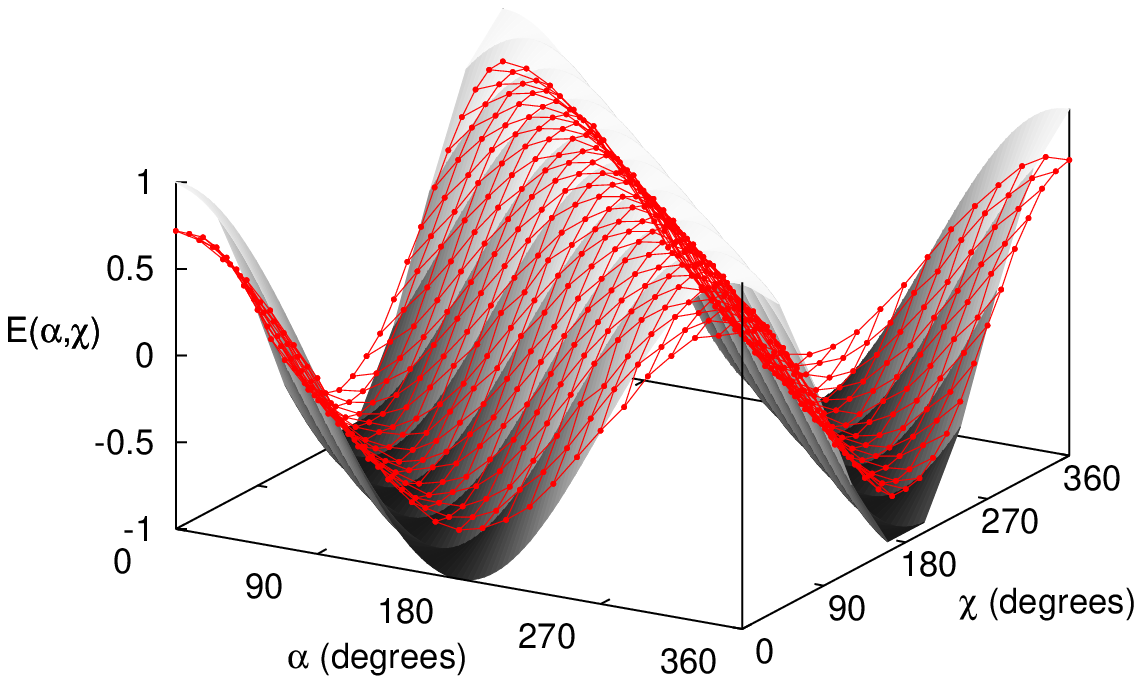}
\caption{%
Left: correlation $E(\alpha,\chi)$ between spin and path degree of freedom as obtained from an event-based simulation of the experiment
depicted in Fig.~\ref{fig5}. Solid surface: $\widehat E(\alpha,\chi)=\cos(\alpha+\chi)$ predicted by quantum theory; circles: simulation data.
The lines connecting the markers are guides to the eye only. Model parameters: reflection percentage of BS0, \ldots, BS3 is 20\% and $\gamma=0.99$.
For each pair $(\alpha,\chi)$, four times 10000 particles were used to determine the four counts $N(\alpha,\chi)$, $N(\alpha+\pi,\chi+\pi)$,
$N(\alpha,\chi+\pi)$ and $N(\alpha+\pi,\chi+\pi)$.
Right: same as figure on the left but $\gamma =0.55$.
}
\label{fig6}
\end{center}
\end{figure}

\subsection{Simulation results}
In Fig.~\ref{fig6}(left) we present simulation results for the correlation $E(\alpha,\chi)$, assuming that the experimental conditions
are very close to ideal and compare them to the quantum theoretical result.

The quantum theoretical result for the correlation $\widehat E_\mathrm{O}(\alpha,\chi)$ is given by~\cite{HASE03}
\begin{eqnarray}
\widehat E_\mathrm{O}(\alpha,\chi)&\equiv&\frac{%
p_\mathrm{O}(\alpha,\chi)+p_\mathrm{O}(\alpha+\pi,\chi+\pi)-p_\mathrm{O}(\alpha+\pi,\chi)-p_\mathrm{O}(\alpha,\chi+\pi)
}{
p_\mathrm{O}(\alpha,\chi)+p_\mathrm{O}(\alpha+\pi,\chi+\pi)+p_\mathrm{O}(\alpha+\pi,\chi)+p_\mathrm{O}(\alpha,\chi+\pi)
}\nonumber \\
&=&\cos(\alpha+\chi)
,
\label{app7}
\end{eqnarray}
where use has been made of the fact that the probability to detect a neutron with spin up in the O-beam is given by
\begin{eqnarray}
p_\mathrm{O}(\alpha,\chi)={\cal T}{\cal R}^2  \left[1+\cos(\alpha+\chi)\right]
,
\label{app6}
\end{eqnarray}
whith $\chi=\chi_0-\chi_1$, and ${\cal R}=1-{\cal T}$ the reflection of the beam splitters (which have been assumed to be identical).
The fact that $\widehat E_\mathrm{O}(\alpha,\chi)=\cos(\alpha+\chi)$ implies that the state of the neutron
cannot be written as a product of the state of the spin and the phase.
In other words, in quantum language, the spin- and phase-degree-of-freedom are entangled~\cite{BASU01,HASE03}.Repeating the calculation for the probability of detecting a neutron in the H-beam shows
that $\widehat E_\mathrm{H}(\alpha,\chi)=0$, independent of the direction of the spin.
If the mu-metal would rotate the spin about the $x$-axis instead of about the $y$-axis, then
we would find $\widehat E_\mathrm{O}(\alpha,\chi)=\cos\alpha\cos\chi$, a typical expression for a quantum system in a product state.

As shown by the markers in Fig.~\ref{fig6} (left), disregarding the small statistical fluctuations,
there is close-to-perfect agreement
between the event-based simulation data for nearly ideal experimental conditions ($\gamma =0.99$ and ${\cal R}=0.2$) and quantum theory.
However, the laboratory experiment suffers from unavoidable imperfections, leading to a reduction and distortion of the interference fringes~\cite{HASE03}.
In the event-based approach it is trivial to incorporate mechanisms for different sources of imperfections by modifying or adding update rules.
However, to reproduce the available data it is sufficient to use the parameter $\gamma$ to control the deviation from the quantum theoretical result.
For instance, for $\gamma=0.55$, ${\cal R}=0.2$ the simulation results for $E(\alpha , \chi )$ are shown in Fig.~\ref{fig6} (right).

In order to quantify the difference between the simulation results, the experimental results and quantum theory
it is customary to form the Bell-CHSH function~\cite{BELL93,CLAU69}
\begin{equation}
\widehat S=\widehat S(\alpha,\chi,\alpha^{\prime},\chi^{\prime})
= \widehat E_\mathrm{O}(\alpha,\chi)
+\widehat E_\mathrm{O}(\alpha,\chi^{\prime})
-\widehat E_\mathrm{O}(\alpha^{\prime},\chi)
+\widehat E_\mathrm{O}(\alpha^{\prime},\chi^{\prime})
,
\label{app8}
\end{equation}
for some set of experimental settings $\alpha$, $\chi$, $\alpha^{\prime}$, and $\chi^{\prime}$.
If the quantum system can be described by a product state, then $|\widehat S|\le2$.
If $\alpha=0$, $\chi=\pi/4$, $\alpha^{\prime}=\pi/2$, and $\chi^{\prime}=\pi/4$, then
$\widehat S\equiv \widehat S_{max}=2\sqrt{2}$, the maximum value allowed by quantum theory~\cite{CIRE80}.

For $\gamma =0.55$, ${\cal R}=0.2$ the simulation results yield
$S_{max}=2.05$, in excellent agreement with the value $2.052\pm 0.010$
obtained in experiment~\cite{HASE03}. For $\gamma=0.67$, ${\cal R}=0.2$ the simulation yields $S_{max}=2.30$, in excellent agreement with the value $2.291\pm 0.008$
obtained in a similar, more recent experiment~\cite{Bartosik2009}.

In conclusion, since experiment shows that $|S|>2$, according to quantum theory it is impossible
to interpret the experimental result in terms of a quantum system in the product state~\cite{BALL03}.
The system must be described by an entangled state.
Meanwhile, the event-based simulation which makes use of classical, Einstein-local and causal event-by-event processes
can reproduce all features of this entangled state.

\subsection{Why are results from quantum theory produced?}
From Ref.~\citen{MICH11a} we know that the event-based model for the beam splitter produces results corresponding to those
of classical wave or quantum theory when applied in interferometry experiments.
Important for this outcome is that the phase difference $\chi$ between the two paths in the interferometer
is constant for a relatively large number of incoming particles.
If, for each incoming neutron, we pick the angle $\chi$ randomly from the same set of predetermined values to
produce Fig.~\ref{fig6}, an event-based simulation
with $\gamma=0.99$ yields (within the usual statistical fluctuations) the correlation
$E(\alpha,\chi)\approx [\cos(\alpha+\chi)]/2$, which does not lead to a violation of the Bell-CHSH inequality (results not shown).
Thus, if the neutron interferometry experiment could be repeated with random choices for the phase shifter $\chi$ for each incident neutron,
and the experimental results would show a significant violation of the Bell-CHSH inequality, then the event-based model that we
have presented here would be ruled out.

\section{Conclusions}
The event-based simulation model provides a
cause-and-effect description of a laboratory single-photon EPRB experiment~\cite{WEIH98,WEIH00}
at a level of detail conventionally overlooked in quantum theoretical descriptions, such as the effect of the choice of the time-window $W$.
Using the same post-selection procedure as the one used in experiment the simulation model can exactly reproduce the results of quantum theory of the EPRB experiment,
namely the single-particle averages and two-particle correlations of the singlet state,
indicating that there is no fundamental obstacle for an EPRB experiment
to produce data that can be described by quantum theory.
However, it is highly unlikely that quantum theory describes the
data of laboratory EPRB experiments which have been performed up to today.~\cite{RAED12}
This suggests that in the real experiment, there may be processes at work which have not been identified yet.

Although the post-selection procedure is essential for the single-photon EPRB experiment to produce
results corresponding to those of its quantum theoretical description, it is absent
in the Bell test experiment with single neutrons~\cite{HASE03} and therefore also in the event-based simulation of it.
As we have demonstrated in this paper, the post-selection procedure is also superfluous in the event-based simulation of
EPRB-type experiments which quantum theory describes by an uncorrelated state, namely experiments II and III.
In experiments II and III the photons leaving the source have orthogonal but definite polarization, and
orthogonal but otherwise random polarization
which is changed into a definite polarization by a polarizer placed between the sourse and the measurement station, respectively.

\section*{Acknowledgement}
We would like to thank K. De Raedt, K. Keimpema, F. Jin, S. Miyashita, S. Yuan, and S. Zhao
for many thoughtful comments and contributions to the work and J. Ralston for making suggestions to improve the manuscript.

%\clearpage
\bibliographystyle{spiebib}   %>>>> makes bibtex use spiebib.bst
\bibliography{../../../all13,../../fisherinfo/manuscript/njp/qtextra}

\begin{thebibliography}{10}

\bibitem{EPR35}
A.~Einstein, A.~Podolsky, and N.~Rosen, ``{Can Quantum-Mechanical Description
  of Physical Reality Be Considered Complete?},'' {\em Phys. Rev.}~{\bf 47},
  pp.~777 -- 780, 1935.

\bibitem{BOHM51}
D.~Bohm, {\em Quantum Theory}, Prentice-Hall, New York, 1951.

\bibitem{CIRE80}
B.~S. Cirel'son, ``Quantum generalizations of {Bell}'s inequality,'' {\em Lett.
  Math. Phys.}~{\bf 4}, pp.~93 -- 100, 1980.

\bibitem{CLAU69}
J.~F. Clauser, M.~A. Horne, A.~Shimony, and R.~A. Holt, ``Proposed experiment
  to test local hidden-variable theories,'' {\em Phys. Rev. Lett.}~{\bf 23},
  pp.~880 -- 884, 1969.

\bibitem{WEIH98}
G.~Weihs, T.~Jennewein, C.~Simon, H.~Weinfurther, and A.~Zeilinger,
  ``{Violation of {Bell}'s Inequality under Strict {Einstein} Locality
  Conditions},'' {\em Phys. Rev. Lett.}~{\bf 81}, pp.~5039 -- 5043, 1998.

\bibitem{WEIH00}
G.~Weihs, {\em {Ein Experiment zum Test der Bellschen Ungleichung unter
  Einsteinscher Lokalit\"at}}.
\newblock PhD thesis, University of Vienna, 2000.
\newblock {\url{http://www.uibk.ac.at/exphys/photonik/people/gwdiss.pdf}}.

\bibitem{HASE03}
Y.~Hasegawa, R.~Loidl, G.~Badurek, M.~Baron, and H.~Rauch, ``{Violation of a
  Bell-like inequality in single-neutron interferometry},'' {\em Nature}~{\bf
  425}, pp.~45 -- 48, 2003.

\bibitem{MICH11a}
K.~{Michielsen}, F.~Jin, and H.~{De Raedt}, ``{Event-based Corpuscular Model
  for Quantum Optics Experiments},'' {\em J. Comp. Theor. Nanosci.}~{\bf 8},
  pp.~1052 -- 1080, 2011.

\bibitem{RAED12a}
H.~{De Raedt} and K.~{Michielsen}, ``{Event-by-event simulation of quantum
  phenomena},'' {\em Ann. Phys. (Berlin)}~{\bf 524}, pp.~393 -- 410, 2012.

\bibitem{RAED12b}
H.~{De Raedt}, F.~Jin, and K.~{Michielsen}, ``{Event-based simulation of
  neutron interferometry experiments},'' {\em Quantum Matter}~{\bf 1}, pp.~1 --
  21, 2012.

\bibitem{ZHAO08}
S.~{Zhao}, H.~{De Raedt}, and K.~Michielsen, ``{Event-by-event simulation model
  of Einstein-Podolosky-Rosen-Bohm experiments},'' {\em Found. Phys.}~{\bf 38},
  pp.~322 -- 347, 2008.

\bibitem{CLAU74}
J.~F. Clauser and M.~A. Horne, ``Experimental and consequences of objective
  local theories,'' {\em Phys. Rev. D}~{\bf 10}, pp.~526 -- 535, 1974.

\bibitem{RAED12}
H.~{De Raedt}, K.~Michielsen, and F.~Jin, ``{Einstein-Podolsky-Rosen-Bohm
  laboratory experiments: Data analysis and simulation},'' {\em AIP Conf.
  Proc.}~{\bf 1424}, pp.~55 -- 66, 2012.

\bibitem{BELL93}
J.~S. Bell, {\em {Speakable and Unspeakable in Quantum Mechanics}}, Cambridge
  University Press, Cambridge, 1993.

\bibitem{AGUE09}
M.~B. {Ag\"uero}, A.~A. Hnilo, M.~G. Kovalsksy, and M.~A. Larotonda, ``{Time
  stamping in EPRB experiments: application on the test of non-ergodic
  theories},'' {\em Eur. Phys. J. D}~{\bf 55}, pp.~705 --709, 2009.

\bibitem{VIST12}
A.~Vistnes and G.~Adenier, ``{There may be more to entangled photon experiments
  than we have appreciated so far},'' {\em AIP Conf. Proc.}~{\bf 1508}(1),
  pp.~326 -- 333, 2012.

\bibitem{ADEN12}
G.~Adenier, ``{Characterization of our source of polarization-entangled
  photons},'' {\em AIP Conf. Proc.}~{\bf 1508}(1), pp.~115 -- 124, 2012.

\bibitem{RAED07b}
K.~{De Raedt}, H.~{De Raedt}, and K.~Michielsen, ``{A computer program to
  simulate Einstein-Podolsky-Rosen-Bohm experiments with photons},'' {\em Comp.
  Phys. Comm.}~{\bf 176}, pp.~642 -- 651, 2007.

\bibitem{KARL09}
K.~{Hess}, K.~{Michielsen}, and H.~{De Raedt}, ``{Possible Experience: from
  Boole to Bell},'' {\em Europhys. Lett.}~{\bf 87}, p.~60007, 2009.

\bibitem{KARL10}
K.~{Hess}, K.~{Michielsen}, and H.~{De Raedt}, ``{Reply to Comment by A.J.
  Leggett and Anupam Garg},'' {\em Europhys. Lett.}~{\bf 91}, p.~40002, 2010.

\bibitem{RAED11a}
H.~{De Raedt}, K.~{Hess}, and K.~{Michielsen}, ``{Extended Boole-Bell
  inequalities applicable to quantum theory},'' {\em J. Comp. Theor.
  Nanosci.}~{\bf 8}, pp.~1011 -- 1039, 2011.

\bibitem{NIEU11}
T.~M. Nieuwenhuizen, ``{Is the Contextuality Loophole Fatal for the Derivation
  of Bell Inequalities?},'' {\em Found. Phys.}~{\bf 41}, pp.~580 -- 591, 2011.

\bibitem{KHRE09}
A.~Y. Khrennikov, {\em {Contextual Approach to Quantum Formalism}}, Springer,
  Berlin, 2009.

\bibitem{KHRE11}
A.~Y. Khrennikov, ``{On the role of probabilistic models in quantum physics:
  Bell's inequality and probabilistic incompatibility},'' {\em J. Comp. Theor.
  Nanosci.}~{\bf 8}, pp.~1006 -- 1010, 2011.

\bibitem{FINE82}
A.~Fine, ``{Some Local Models for Correlation Experiments},'' {\em
  Synthese}~{\bf 50}, pp.~279 -- 294, 1982.

\bibitem{PASC86}
S.~Pascazio, ``{Time and Bell-type Inequalities},'' {\em Phys. Lett. A}~{\bf
  118}, pp.~47 -- 53, 1986.

\bibitem{LARS04}
J.-{\AA}. Larsson and R.~D. Gill, ``{Bell}'s inequality and the
  coincidence-time loophole,'' {\em Europhys. Lett.}~{\bf 67}, pp.~707 -- 713,
  2004.

\bibitem{HNIL02}
A.~Hnilo, A.~Peuriot, and G.~Santiago, ``{Local Realistic Models Tested by the
  EPRB experiment with Variable Analyzers},'' {\em Found. Phys. Lett.}~{\bf
  15}, pp.~359 -- 371, 2002.

\bibitem{RAUC00}
H.~Rauch and S.~A. Werner, {\em Neutron Interferometry: Lessons in Experimental
  Quantum Mechanics}, Clarendon, London, 2000.

\bibitem{KROU00}
G.~Kroupa, G.~Bruckner, O.~Bolik, M.~Zawisky, M.~Hainbuchner, G.~Badurek, R.~J.
  Buchelt, A.~Schricker, and H.~Rauch, ``Basic features of the upgraded s18
  neutron interferometer set-up at ill,'' {\em Nucl. Instrum. Methods Phys.
  Res. A.}~{\bf 440}, pp.~604 -- 608, 2000.

\bibitem{BASU01}
S.~Basu, S.~Bandyopadhyay, G.~Kar, and D.~Home, ``Bell's inequality for a
  single spin-1/2 particle and quantum contextuality,'' {\em Phys. Lett.
  A}~{\bf 279}(5 -- 6), pp.~281 -- 286, 2001.

\bibitem{Bartosik2009}
H.~Bartosik, J.~Klepp, C.~Schmitzer, S.~Sponar, A.~Cabello, H.~Rauch, and
  Y.~Hasegawa, ``Experimental test of quantum contextuality in neutron
  interferometry,'' {\em Phys. Rev. Lett.}~{\bf 103}, p.~040403, 2009.

\bibitem{BALL03}
L.~E. Ballentine, {\em {Quantum Mechanics: A Modern Development}}, World
  Scientific, Singapore, 2003.

\end{thebibliography}

\end{document}